\documentstyle[12pt]{article}

\def\shiftdown#1{#1\llap{\lower.04ex\hbox{#1}}}

\begin{document}

\vspace*{0.3cm}

\begin{center}

{\bf {\large Conformal transformations and doubling of the particle states
 }}

\end{center}

\begin{center}
{\large 

{\it A.\ I.\ Machavariani$^{a,b}$} }

{\em $^{a}$ Joint\ Institute\ for\ Nuclear\ Research,\ Moscow
Region 141980 Dubna,\ Russia}\\[0pt]

{\em $^{b}$ High Energy Physics Institute of Tbilisi State University,
University Str. 9 }\\[0pt]

\end{center}

\vspace{0.5cm} \medskip

\begin{abstract}

The 6D and 5D representations of the four-dimensional (4D)   
interacting fields in the Heisenberg picture
and the corresponding  equations of motion are 
studied  using equivalence of
the conformal transformations of the four-momentum $q_{\mu}$
($q'_{\mu}=q_{\mu}+h_{\mu}$, $q'_{\mu}=\Lambda^{\nu}_{\mu}q_{\nu}$, 
$q'_{\mu}=\lambda q_{\mu}$ and  $q'_{\mu}=-M^2q_{\mu}/q^2$)
and the corresponding rotations  on the 6D cone
$\kappa_A\kappa^A=0$ $(A=\mu;5,6\equiv 0,1,2,3;5,6)$,
where $q_{\mu}=M\ \kappa_{\mu}/(\kappa_{5}+\kappa_{6})$ and 
$M$ is the scale parameter.
The 4D reduction of the 6D fields on the cone $\kappa_A\kappa^A=0$
is unambiguously fulfilled by the intermediate 5D projection 
into two 5D hyperboloids  $q_{\mu}q^{\mu}+ q_5^2= M^2$ and
$q_{\mu}q^{\mu}- q_5^2=- M^2$ in order to cover the whole  domains
$-\infty<q_{\mu}q^{\mu}<\infty$ and $q_5^2\ge 0$. 
The resulting  5D and  4D fields in the coordinate space
consist of two parts $\varphi_1(x,x_5)$, $\varphi_2(x,x_5)$ and 
$\Phi_1(x)=\varphi_1(x,x_5=0)$, $\Phi_2(x)=\varphi_2(x,x_5=0)$,
where the Fourier conjugate of
$\varphi_1(x,x_5)$ and $\varphi_2(x,x_5)$ are defined on the 
hyperboloids   $q_{\mu}q^{\mu}+ q_5^2= M^2$ and
$q_{\mu}q^{\mu}- q_5^2=- M^2$ respectively. Consequently,
the 4D reduction of the  6D fields 
generate two kinds  of the  5D and 4D  fields
$\varphi_{\pm}=\varphi_1\pm\varphi_2$ and 
$\varphi_{\pm}(x,x_5=0)=\Phi_{\pm}(x)=\Phi_1(x)\pm\Phi_2(x)$ 
with the same quantum numbers but with the different 
masses  and the sources.  
This  doubling of the 4D fields 
$\Phi_{\pm}=\Phi_1\pm \Phi_2$ can be applied for  
unified description of the interacting 
electron and muon fields, $\pi$ and $\pi(1300)$-mesons, $N$
and $N(1440)$-nucleons and other particles with the same 
quantum numbers but different masses and interactions.

 

\end{abstract}

\newpage

\begin{center}
{\bf{ Introduction}}
\end{center}

\vspace{0.5cm}

The 5D extension of the 4D relativistic theories is the fruitful method
that has a long history.  
The Kaluza-Klein theory  and their generalizations for the gauge 
transformations \cite{Appelquist,Bailin,Wesson,Cian}
allow to  unify the electromagnetic and gravitation theories.
In the traditional Kaluza-Klein theory all partial derivatives 
with respect to fifth coordinates have been equated to zero
and the extra spatial dimension was compacted to a small size circle.
The rigorous mathematical approach for the $N+1$ and $N$ dimensional
manifolds (see ch. 2.2 in \cite{Wesson})
allow to embed the 4D equation  of motion  with the sources into the 5D 
equation without sources. In the recent 5D field theoretical 
formulations \cite{Fukuyama,Fujimoto} the extra fifth dimension is required
to solve the problems of the renormalizable $SO(10)$ 
grand unification theories with the breakdown of the gauge coupling.
In this approach the fifth dimension enable 
 to reproduce  the fermion generations, quark mass hierarchy, 
flavor mixing and Cabbibbo-Kabayashi-Maskawa  matrices
and it is argued,  that in virtue of the  
no go theorems it is not possible to achieve these results in the 4D space.
 Other kind of the 5D  relativistic field theories
were performed within the invariant time method   
\cite{IZ,Fanchi,Land},
where the fifth coordinate is  the proper time
$x_5^2=x_o^2-{\bf x}^2$. 
In these theories $x_5$ is an auxiliary variable and the sought
4D wave functions and fields  are reproduced through
the 5D wave functions and fields via the boundary conditions
for $x_5=0$ or $x_5=\sqrt{t^2-{\bf x}^2}$ and
the evolution over  the fifth coordinates   were often described 
through the equation for the first derivatives of the scalar and 
fermion fields  $i\partial \phi/\partial x_5$ and  
$i\partial \psi/\partial x_5$.



The general scheme for the 5D extensions of the 4D relativistic theories 
and the 4D reductions of the 5D relativistic formulations
presents the conformal group of  the 4D transformations that can
 be unambiguously represent through the rotations on the 6D cone.
In particular,  
the conformal transformations of the four coordinate $x_{\mu}$ 
consists  of the following independent  motions
$x'_{\mu}=x_{\mu}+a_{\mu}$,
$x'_{\mu}=\Lambda^{\nu}_{\mu}x_{\nu}$, 
$x'_{\mu}=\lambda x_{\mu}$ and  $x'_{\mu}=
-\ell^2x_{\mu}/x^2$ which  can be  performed through the rotations
on the 6D cone 
$\xi_A\xi^A\equiv \xi_{\mu}\xi^{\mu}+
\xi_{5}\xi^{5}-\xi_{6}\xi^{6}=0$ \cite{Dir},  
where $A=0,1,2,3;5,6\equiv \mu;5,6$, 
$x_{\mu}=\xi_{\mu}/\xi_+$,\ \ \ $\xi_{\pm}=(\xi_5\pm\xi_6)/\ell$
and $\ell$ is the dimension parameter.
The one-to-one relationship between  the 4D conformal transformations
and 6D rotations allow to construct the one-to-one relationship between
an interacting 4D Heisenberg field $\Phi(x)$ and 
the corresponding 6D field
$\varsigma(\xi)\equiv 
\varsigma(\xi_0,\xi_1,\xi_2,\xi_3;\xi_5,\xi_6)$ as
$\Phi(x)=\Bigl[ \varsigma(x,\xi_+,\xi_-)\Bigr]_{\xi_A\xi^A=0}$
with the fixed scale parameter $\xi_+$ and   
${\xi_-}/\xi_+=x^2/\ell^2$ \cite{Dir}-\cite{Ca2}.
The other  4D reduction of $\varsigma(\xi)$ was used 
in the manifestly  conformal invariant formulation 
\cite{M1}-\cite{Todorov},
 where the homogeneity of the conformal invariant 6D fields
$\Bigl[\varsigma(\xi)\Bigr]_{\xi_A\xi^A=0}=\Bigl[\varsigma(x,\xi_+,\xi_A\xi^A=0)$
over the scale variable $\xi_+$ is required, i.e.
$\varsigma(x,\xi_+,\xi_A\xi^A=0)=
{\xi_+}^{d}{ \phi}(x,\xi_A\xi^A=0)$ and the 4D conformal invariant field is
$\Phi(x)\equiv { \phi}(x,\xi_A\xi^A=0)$.

Location of $\varsigma(\xi)$ on the 6D cone $\xi_A\xi^A=0$
impose additional condition by conformal transformations.
For instance, the Fourier transformation of 
an arbitrary field $\varsigma(\xi)$ on the 6D cone produces the following 
condition
$$\biggl({ {\partial^2}\over{\partial {\kappa}^{\mu}\partial {\kappa}_{\mu}}}
+{ {\partial^2}\over{\partial {\kappa}^{5}\partial {\kappa}_{5}}}
-{ {\partial^2}\over{\partial {\kappa}^{6}\partial {\kappa}_{6}}}\biggr)
\int d^6\xi e^{i\kappa_A\xi^A}\delta\Bigl(\xi_o^2-\xi_1^2-\xi_2^2-\xi_3^2
+\xi_5^2-\xi_6^2\Bigr)\varsigma(\xi)=0,\eqno(I.1a)$$
where $\kappa_A$ are Fourier conjugate to $\xi_A$ and for derivation of (I.1a) 
the condition  $(\xi_A\xi^A)\delta(\xi_A\xi^A)=0$ was used.
The intermediate 5D projection of the 6D fields and the condition (I.1a)
determine the corresponding 5D fields and the 5D condition.

In order to obtain the 5D and 6D extensions of the 4D equations of motion
it is convenient to consider the conformal transformations
of the four momentum $q_{\mu}$  
($q'_{\mu}=q_{\mu}+h_{\mu}$, $q'_{\mu}=\Lambda^{\nu}_{\mu}q_{\nu}$, 
$q'_{\mu}=\lambda q_{\mu}$ and  $q'_{\mu}=-M^2q_{\mu}/q^2$).
The 6D representation of the conformal transformations
for the independent four components of the momentum $q_{\mu}$ 
are similar with the conformal transformations in the  coordinate space.
The principal difference between the conformal transformations
in the coordinate and momentum space is in the translation. 
In the next section it is shown that the translation of the four momentum
$q'_{\mu}=q_{\mu}+h_{\mu}$
for the Fourier conjugate of $\Phi(x)$ produces the gauge transformation
 $\Phi'(x)=e^{ih_{\mu}x^{\mu}}\Phi(x)$.
According to the   Dirac geometrical model \cite{Dir}-\cite{BK}, 
each of the conformal transformations in the momentum space
is unambiguously determined 
via the appropriate 6D rotation with the invariant 6D form

$$\kappa_A\kappa^A\equiv\kappa_{\mu}\kappa^{\mu}+\kappa_5^2-\kappa_6^2=0,
\eqno(I.2a)$$

where the four momentum
$q_{\mu}$ ($\mu=0,1,2,3$) is defined as
$q_{\mu}= \kappa_{\mu}/\kappa_+ $ and
$M$ is a scale parameter.
The 6D cone (I.2a) and the corresponding surface 

$$q_{\mu}q^{\mu}+M^2
{{\kappa_-}\over{\kappa_+}}=0,\ \ \ with \ \ \ 
 \kappa_{\pm}= {{\kappa_5\pm\kappa_6}\over M}
\eqno(I.2b)$$
are invariant  under  any combination of the
conformal transformations of $q_{\mu}$.

In analogy with (I.1a) the conformal transformations  of a 4D field  
$\Phi(x)$ in the momentum space 
can be performed via  the 6D rotations of the
 corresponding 6D field operator $\varsigma(\kappa)$ which  is
embedded into the cone (I.2a). Therefore,  
location  of $\varsigma(\kappa)$  on the same 6D cone (I.2a) 
before and after the conformal transformations in the momentum space
impose the condition

$$\biggl({ {\partial^2}\over{\partial {\xi}^{\mu}\partial {\xi}_{\mu}}}
+{ {\partial^2}\over{\partial {\xi}^{5}\partial {\xi}_{5}}}-
{ {\partial^2}\over{\partial {\xi}^{6}\partial {\xi}_{6}}}\biggr)
\int {{d^6\kappa}\over{(2\pi)^4}} 
e^{i\kappa_A\xi^A}\delta\Bigl(\kappa_{\mu}\kappa^{\mu}
+\kappa_5^2-\kappa_6^2\Bigr)\varsigma(\kappa)=0.\eqno(I.1b)$$

This paper deals with  consistency of the usual 4D equations of motion for 
4D interacting Heisenberg field $\Phi(x)$
and boundary conditions and constrains for the 5D and 6D representations of $\Phi(x)$
which follows from the conformal group of the transformations in the momentum space.
Two particular features 
generate the special interest to the conformal transformations in
the momentum space\cite{BR}. First, the observables of the
particle interactions, like the cross sections and polarizations 
are determined in the momentum space.
Secondly, the accuracy of the measurement of the particle coordinates
is in principle restricted by the Compton length of this
particle. Moreover, determination of the coordinate
of the conformal invariant   massless particles 
produces additional essential troubles 
(see  \cite{BaR} ch. 20 and \cite{Bacr}). 
The conformal transformations of the fields 
and  the corresponding equations of motion 
in the momentum space were considered in ref.
\cite{Kas,Ca2,BR,Budinich}, where  the conformal
transformations were performed in the configuration space 
and  followed relations  in the momentum space were obtained 
using the Fourier transformation.

The 4D reduction of the 6D operators 
$\varsigma(\kappa)\equiv\varsigma(q,\kappa_+,\kappa_-)$ 
generates the 5D operators as the intermediate 5D projections.
There  exists only two 5D  De Sitter spaces with the constant 
curvature which
have the  invariant forms $q_{\mu}q^{\mu}\pm q_5^2\mp M^2=0$
($q_5^2\ge 0$) of the 
$O(2,3)$ and $O(1,4)$ rotational groups \cite{BK,Todorov,K2}.
The single 5D hyperboloid  is not enough for 
reproduction of the whole values of $-\infty<q^2<\infty$. 
Therefore, we use
the domains from the  
both 5D hyperboloids which are connected by
inversion $q'_{\mu}=-M^2q_{\mu}/q^2$.
Thus for the intermediate 
5D projections of the 6D cone (I.2a) we shall use the
  invariant forms
$$q_{\mu}q^{\mu} + q_5^2= M^2\ \ \ with\ \ \  
{q_5^2 \over M^2}= 
{ {\kappa_{-}} \over{\kappa_{+} } } +1,\ \ \ and\ \ q_5^2\ge 0
\eqno(I.3a)$$

$$q_{\mu}q^{\mu} - q_5^2= -M^2\ \ with\ \ \ 
{q_5^2 \over M^2}=-{ {\kappa_{-}} \over{\kappa_{+} } } +1
\ \ \ and\ \ q_5^2\ge 0.
\eqno(I.3b)$$

In (I.3a,b) the fifth variable $q_5^2$  is positive $0\le q_5^2\lq \infty$.   
Consequently, for the positive  
$q^2\equiv q_{\mu}q^{\mu}\ge 0$  the corresponding four-momenta $q_{\mu}$
are distributed between the domains $0\leq q^2\leq M^2$ and 
$q^2> M^2$ on the hyperboloids (I.3a) and (I.3b) respectively.
These domains are connected by inversion $q'_{\mu}=-M^2 q_{\mu}/q^2$.
The values of the $q^2$  on 
the hyperboloids (I.3a) and (I.3b) are also connected through the reflection 
$q^2\longleftrightarrow -q^2$.
Therefore, for the negative $q^2<0$ one has $-M^2< q^2<0$ 
on the hyperboloid (I.3b) and 
$ -\infty<q^2<- M^2$ is on the hyperboloid (I.3a).
More detailed  the distributions of $q^2$ on the hyperboloids
(I.3a) and (I.3b) are listed in Table 1 of Section 2.


In order to determine the 5D and 4D  projections of the 
6D fields  $\delta(\kappa_A\kappa^A)\varsigma(\kappa)$ (I.2a) 
we  shall  introduce the following 5D fields
$\varphi_{1}(x,x_5)$ and $\varphi_{2}(x,x_5)$
$$\varphi_{1}(x,x_5)=\int {{d^4q}\over{(2\pi)^4}}dq_5^2e^{-iqx-iq^5x_5}
\delta(q^2+q_5^2-M^2)
\Bigl[\theta(q^2)\theta(M^2-q^2)+\theta(-q^2)\theta(-M^2-q^2)]
\Bigr]\phi(q,q_5^2),\eqno(I.4a)$$

$$\varphi_{2}(x,x_5)=
\int {{d^4q}\over{(2\pi)^4}}dq_5^2e^{-iqx-iq^5x_5}
\delta(q^2-q_5^2+M^2)
\Bigl[\theta(q^2)\theta(-M^2+q^2)+\theta(-q^2)\theta(M^2+q^2)\Bigr]
\phi(q,q_5^2).\eqno(I.4b)$$

where

$$\phi(q,q_5^2)={{M^2}\over 2}\int \kappa_+^3d\kappa_+ \theta(\kappa_+)
 \varsigma(q,q_5^2,\kappa_+).\eqno(I.5)$$
For  the sake of simplicity 
the scale variable $\kappa_+$ (I.2b) is taken in positive i.e.
 $\phi(\kappa)=\theta(\kappa_+)\phi(\kappa)$, where 
$\theta(\kappa_+)=1$ for $\kappa_+>0$ and $\theta(\kappa_+)=0$ for $\kappa_+<0$.

The Fourier conjugate of $\varphi_{1}(x,x_5)$  and $\varphi_{2}(x,x_5)$
are located into hyperboloids (I.3a) and 
(I.3b) respectively. Therefore they satisfy the conditions

$$\biggl({{\partial^2}\over{\partial x^{\mu}\partial x_{\mu}}}+
{{\partial^2}\over{\partial x^5\partial
x_5}}+M^2\biggr)\varphi_1(x,x_5)=0,\ \ \ \ \ \ 
\ \ \ \biggl({{\partial^2}\over{\partial x^{\mu}\partial x_{\mu}}}-
{{\partial^2}\over{\partial x^5\partial
x_5}}-M^2\biggr)\varphi_2(x,x_5)=0. \eqno(I.6)$$

The fields $\varphi_{1}$ (I.4a) and 
$\varphi_{2}$ (I.4b)  produce  two independent 5D fields  

$$\varphi_+(x,x_5)=\varphi_1(x,x_5)+\varphi_2(x,x_5);
\ \ \ \varphi_-(x,x_5)=\varphi_1(x,x_5)-\varphi_2(x,x_5),\eqno(I.7)$$
which Fourier conjugate  are defined in the whole domains
$(-\infty<q_{\mu}<+\infty)$ and $(-\infty<q^2<+\infty)$.

The usual boundary condition for the 5D fields $\varphi_{\pm}(x,x_5)$
at $x_5=0$ allows to get the 4D fields $\Phi_{\pm}(x)$

$$\Phi_{\pm}(x)=\varphi_{\pm}(x,x_5=0).\eqno(I.8)$$

According to (I.4a,b), (I.7) and (I.8) the Fourier conjugate of $\Phi_{\pm}(x)$ 

$$\Phi_{\pm}(q)=\int d^4x e^{iqx}\Phi_{\pm}(x)\eqno(I.9)$$

have the following structure

$$\Phi_{\pm}(q)=\sum_{N=I,III}\Phi_{N}(q)\pm\sum_{N=II,IV}\Phi_{N}(q);\eqno(I,10)$$

where

$$\Phi_{I}(q)= \theta(q^2)\theta(M^2-q^2)\phi(q,q^2_5=M^2-q^2);\ \ \
\Phi_{II}(q)= \theta(q^2)\theta(-M^2+q^2)\phi(q,q^2_5=M^2+q^2);$$
$$\Phi_{III}(q)= \theta(-q^2)\theta(-M^2-q^2)\phi(q,q^2_5=M^2-q^2);\ \ \
\Phi_{IV}(q)= \theta(-q^2)\theta(M^2+q^2)\phi(q,q^2_5=M^2+q^2),\eqno(I.11)$$
where  the lower index $I,\ II,\ III$ and $IV$ of $\Phi(q)$ corresponds to 
 the domains of $q^2$ which are listed in Table 1 of Section 2.
The details of the relationship between the 4D and 5D scalar field are given in 
in Section 3.  

The equations (I.4a,b)-(I.10) presents the relationship between the
 4D, 5D and  
6D fields  $\delta(\kappa_A\kappa^A)\varsigma(\kappa)$.
The 4D interacting Heisenberg fields
$\Phi_{\pm}(x)$  and their 6D representations 
$\delta(\kappa_A\kappa^A)\varsigma(\kappa)$ are not invariant under the conformal 
transformations. Nevertheless, location of  $\delta(\kappa_A\kappa^A)\varsigma(\kappa)$
on the 6D cone (I.2a) and the corresponding 
location of the Fourier conjugate of 
$\varphi_{\pm}(x,x_5)$ on the hyperboloids (I.3a,b) produce the 
conditions (I.1b) and (I.6) for any 6D field and its 5D projections.
The consistency of these conditions with the equation of motion
for $\varphi_{\pm}(x,x_5)$ and $\Phi_{\pm}(x)$ and the boundary conditions
(I.8) are considered in Sect. 4 and 5.

The intermediate projection of the
6D field $\delta(\kappa_A\kappa^A)\varsigma(\kappa)$ on the
invariant forms  (I.3a,b) of the $O(2,3)$ and $O(1,4)$  
subgroups of the conformal  group $O(2,4)$
need to introduce the two independent
 5D  fields  $\varphi_+$ and $\varphi_-$  (I.7)
which are constructed from the same parts $\varphi_1$ and $\varphi_2$ (I.4a,b).
The invariant forms (I.3a,b) of the $O(2,3)$ and $O(1,4)$
form the definition area of   $\varphi_1$ and $\varphi_2$
and correspondingly of the field $\Phi_N(q)$ in (I.10)  
and  (I.11).  
The  4D fields $\Phi_{+}$ and $\Phi_{-}$ (I.10)
have the same quantum numbers. 
But $\Phi_+$ and
$\Phi_-$ can have the different  masses and the different sources.
Other details of the inversion and related constructions
of the  4D fields are given in the next Section.

The reduction formulas (I.5)  of the 6D field
$\varsigma(\kappa)$ on the cone (I.2a)  
$\varsigma(\kappa)\equiv\varsigma(\kappa_{\mu},\kappa_+,\kappa_A\kappa^A\ne 0)$
 $\equiv\varsigma(q,q_5^2,\kappa_+)$ differ from the reduction formula
in the manifestly covariant formulation \cite{M1}-\cite{Todorov},
where the homogeneity of  
$\varsigma(\kappa)=\phi_(q_{\mu},\kappa_+,\kappa_A\kappa^A=0)$ over the
scale variable $\kappa_+$ is required, i.e.
$\phi(q_{\mu},\kappa_+,\kappa_A\kappa^A=0)=
{\kappa_+}^d{ \Phi}(q_{\mu}),$
where $d$ is the scale dimension of $\phi(\kappa)$.
In order to reproduce this property
in the present approach one can use an additional 
condition in (I.5)

$$\varsigma(q,q_5^2,\kappa_+)=\delta(\kappa_+-{\cal M})\phi(q,q_5^2,{\cal M});\
\ \ \ \ \ or \ \ \ \ \ \ \  
\varsigma(q,q_5^2,\kappa_+)=\theta(\kappa_+-{\cal M})\phi(q,q_5^2,{\cal M})
\eqno(I.12)$$
where ${\cal M}$ is a fixed  scale parameter. 


The 5D  quantum  field theory with the invariant forms 
$q_{\mu}q^{\mu} + q_5^2= M^2$ or $q_{\mu}q^{\mu} - q_5^2= -M^2$ separately 
was firstly studied in refs.  \cite{K1,K2,K3},
where  $M$ was interpreted as the 
fundamental (maximal) mass
and its inverse $1/M$ as the fundamental (minimal) 
length \cite{Heisenberg,Markov}.
The conformal transformations in the momentum space for the 
complete fields $\varphi_+=\varphi_1+\varphi_2$
was suggested in \cite{prepr}, where $M$ is determined
via $m_{\pi}$ and $m_{Higgs}$ according to the
chiral symmetry breaking mechanism within the 5D chiral models.
In the present paper  is studied coupling between
the 5D and 4D equations of motion for 
the fields $\varphi_+$, $\Phi_+$ and 
$\varphi_-$, $\Phi_-$. Besides the present paper contains
more general and self-consistent formulation of
 translations and inversions in the
 4D momentum space for the 4D charged and neutral fields.

 In Section 1  the conformal transformations 
in the 4D momentum space and the corresponding  
transformations of the interacting   
fields $\Phi(x)$ are considered. 
The domains of the variables   $q^2=q_{\mu}q^{\mu}$ and $q^{2}_5$ 
(I.3a,b) are determined in Sect. 2. The 4D reduction 
of the 5D fields and
the related projections and convolution formulas are given in Sect. 3.
In Sections 4 and 5 the 4D and 5D  equations of motion for the scalar 
fields and the corresponding  constrains for $x_5$
are considered.
Sections 6 is devoted to the 5D and 4D
Lagrangians. The 4D and 5D equations of motion for the fermion
fields  with the electromagnetic interaction and the  constrains
for the fifth coordinates $x_5$ are considered in Sect. 7 and 8.
In Section 9 the 5D extension  of the
standard $SU(2)\times U(1)$ model for the
electron and muon fields is given
as an example of the suggested doubling for the fermion
states. Besides in this Section 
is shortly discussed  consistency of the present scheme and 
the purely 5D models \cite{Fukuyama,Fujimoto} of the grand unification theory. 
The  generalized translations in the momentum space 
as the gauge transformations are considered in Sect 10.
The summary is given in Sect. 11.

\vspace*{0.15cm}

\begin{center}
{\bf{1. \ Conformal transformations 
 in the 4D momentum space.}}
\end{center}

\vspace{0.15cm}

Conformal transformations of the four-momentum 
 $q_{\mu}$ $(\mu=0,1,2,3)$  consists of 

$translations\hfill{q_{\mu}\longrightarrow q_{\mu}'=q_{\mu}+h_{\mu}
,\ \ \ \ \ (1.1a)}$

$rotations\hfill{q_{\mu}\longrightarrow
q_{\mu}'=\Lambda_{\mu}^{\nu}q_{\nu},\ \ \ \ \ (1.1b)}$

$dilatation\hfill{
q_{\mu}\longrightarrow q_{\mu}'=e^{\lambda} \
q_{\mu},\ \ \ \ \ (1.1c)}$

$and \ inversion\hfill{
q_{\mu}\longrightarrow q_{\mu}'=-M^2 q_{\mu}/q^2,\ \ \ \ \ (1.1d)}$
 
where  $M$ is a mass parameter
that  insures the correct dimension of  $q_{\mu}$.
Translations and inversions form   

$special\ conformal\ transformation\hfill{
q_{\mu}\longrightarrow q_{\mu}'=
\frac {\Large {q_{\mu}-{\hbar}_{\mu} q^2/ M^2} }
{\Large {1-2q_{\nu}{\hbar}^{\nu}/M^2+{\hbar}^2q^2/M^4}} .\ \ \ \ \ (1.1e)}$
\par
$q_{\mu}$ in (1.1a)-(1.1e) is off mass shell, i.e. $q_o$ 
is an independent variable  and $q_o\ne\sqrt{{\bf q}^2+m^2}$. 


According to the Dirac geometrical model \cite{Dir},
transformations (1.1a)-(1.1e) are equivalent to the 
rotations on the 
6D cone $\kappa^2\equiv\kappa_A\kappa^A=0$ (I.2a)
with  the metric tensor $g_{AB}=diag(+1,-1,-1,-1,+1,-1)$.
In particular, translation (1.1a) and the special conformal 
translation
are generated by the combinations of the 
rotations in the planes $(\mu,5)$ and $(\mu,6)$, dilatation is obtained
via the rotation in the plane (5,6) and inversion follows from 
transposition of the of the variables $\kappa'_5=\kappa_5$ and 
$\kappa_6'=-\kappa_6$   

$$translation:\ \ \ \ \ \ \ \ \ \ \ \ \ \ 
 \kappa_{\mu}'=\kappa_{\mu}+h_{\mu}\kappa_+;\ \ \ \ \ 
\kappa_+'=\kappa_+;\ \ \ \ \ 
\kappa_-'=-{{2h_{\mu}\kappa^{\mu}}\over{\kappa_+}}-
{{\kappa_{\mu}\kappa^{\mu}}\over{\kappa_+}}\eqno(1.2a)$$
$$rotation:\ \ \ \ \ \ \ \ \ \ \ \ \ \ \ \ \ 
\ \ \ \ \ \ \ \ \ \ \ \ \ \ \ \ \ 
\kappa_{\mu}'=\Lambda_{\mu}^{\nu}\kappa_{\nu};\ \ \ \ \
\kappa_+'=\kappa_+;\ \ \ \ \ \kappa_-'=\kappa_-,\eqno(1.2b)$$
$$dilatation:\ \ \ \ \ \ \ \ \ \ \ \ \ \ \ \ \ \ \ \
\ \ \ \ \ \ \ \ \ \ \ \ \ \ \ \ \ 
\kappa_{\mu}'=\kappa_{\mu};\ \ \ \ \ 
\kappa_{+}'=e^{-\lambda}\kappa_{+};\ \ \ \ \
\kappa_{-}'=e^{\lambda}\kappa_{-},\eqno(1.2c)$$ 
$$inversion:\ \ \ \ \ \ \ \ \ \ \ \ \ \ \ \ \ \ \ \
\ \ \ \ \ \ \ \ \ \ \ \ \ \ \ \ \ 
\kappa_{\mu}'=\kappa_{\mu};\ \ \ \ \
\kappa_{+}'=\kappa_{-};\ \ \ \ \ 
\kappa_{-}'=\kappa_{+},\eqno(1.2d)$$

where
$$q_{\mu}={ {\kappa_{\mu}}\over{\kappa_{+} }};\ \ \ \ \ \ \ \ \ \ \ \ \ \ \ \ \  
\kappa_{\pm}={{\kappa_{5}\pm\kappa_{6}}\over M};\ \ \ \ \ \ \ \ \  
\mu=0,1,2,3.\eqno(1.3)$$

The equivalence of the 4D and 6D conformal transformations 
implies that  translation, rotation, dilatation and inversion   
of the 4D field $\Phi(q)$ (I.9)
are unambiguously (isomorphic)
determined through the corresponding 
6D rotations of the 6D field 
$\varsigma(\kappa)\equiv
\varsigma(\kappa_{\mu};\kappa_+,\kappa_-)$
$$\Phi(q_{\mu}'=q_{\mu}+h_{\mu})\Longleftrightarrow
\varsigma(\kappa_{\mu}=\kappa_{\mu}+h_{\mu}\kappa_+,\kappa_ +,
\kappa_-'=\kappa_--\frac{2h_{\mu}\kappa^{\mu}}{\kappa_+}-
\frac{\kappa_{\mu}\kappa^{\mu}}{\kappa_+})
\eqno(1.4a)$$,
$$\Phi(q_{\mu}'=\Lambda_{\mu}^{\nu}q_{\nu})\Longleftrightarrow 
\varsigma(\kappa_{\mu}'=\Lambda_{\mu}^{\nu}\kappa_{\nu},\kappa_+,\kappa_-)
\eqno(1.4b)$$
 $$\Phi(q_{\mu}'={\lambda} q_{\nu})\Longleftrightarrow
\varsigma(\kappa_{\mu},e^{-\lambda} \kappa_{+},e^{\lambda}\kappa_{-})
\eqno(1.4c)$$
$$ \Phi(q_{\mu}'=-q_{\mu}/q^2)\Longleftrightarrow
\varsigma(\kappa_{\mu},\kappa_ -,\kappa_+)\eqno(1.4d)$$

These relationships between the 4D and 6D operators $\Phi(q)$ and 
$\varsigma(\kappa)$ is achieved in (I.1b) through 
$\delta(\kappa_A\kappa^A)$.

An interacting scalar field  $\Phi(x)$ 
is usually decomposed in the
positive and in the negative frequency parts in the 3D Fock space

$$\Phi(x)=\int {{d^3 p}\over{(2\pi)^3 2\omega_{{\bf p}}} }
\Bigl[
a_{{\bf p}}(x_0)e^{-ipx}+{b^+}_{{\bf p}}(x_0)e^{ipx} \Bigr];
\ \ \ p_o\equiv \omega_{{\bf p}}
=\sqrt{ {\bf p}^2+m^2},\eqno(1.5)$$
where in the asymptotic regions
 $a_{{\bf p}}(x_0)$ and ${b^+}_{{\bf p}}(x_0)$ 
transforms into particle
 (antiparticle) annihilation (creation) operators.
On the other hand, $\Phi(x)$  can be decomposed in the 4D momentum 
space as

$$\Phi(x)=\int {{d^4 q}\over{(2\pi)^4  } }
\Bigl[{\Phi^{(+)}} (q)e^{-iqx}+
{\Phi^{(-)}}^+ (q)e^{iqx}\Bigr];\ \ \
 or\ \ \ \Phi(x)=\int {{d^4 q}\over{(2\pi)^4  } }
\Phi(q)e^{-iqx}.\eqno(1.6a)$$
where
$$\Phi(q)=\Phi^{(+)}(q)+{\Phi^{(-)}}^+(-q)\eqno(1.6b)$$

Comparison of  (1.5) and (1.6a) gives 

$$ {{e^{-i\omega_{\bf p}x_o}}\over{2\omega_{\bf p}}}
a_{\bf p}(x_o)=
\int {{dq_o}\over{2\pi}} {\Phi^{(+)}} (q_o,{\bf p}) e^{-iq_ox_o}
\eqno(1.7a)$$
 and
$${{e^{i\omega_{\bf p}x_o}}\over{ 2\omega_{\bf p}} }
{b^+}_{\bf p}(x_0)=
\int {{dq_o}\over{2\pi}} {\Phi^{(-)}}^+(q_o,{\bf p})e^{iq_ox_o}.
\eqno(1.7b)$$

The field operators $a_{{\bf p}}(x_0)$ and ${b^+}_{{\bf p}}(x_o)$
are simply determined via the corresponding source operator
$\partial a_{{\bf p}}(x_0) / \partial{ x_{0}}=i\int d^3x e^{ipx}
j(x)$,
where
$\Bigl({{\partial^2} /{\partial{ x_{\mu}}\partial{x^{\mu} }
}}+m^2\Bigr) \Phi(x)=j(x)$. Moreover,
these operators determine the transition  ${\cal S}$-matrix 

$${\cal S}_{mn}\equiv
<out;{\bf p'}_{1},...,{\bf p'}_{m}| {\bf
p}_{1},...,{\bf p}_{n};in>=
\prod_{i=1}^m\Bigl[ \int d{{x^0}'}_i {{d}\over{d {{x'}^0}_i}}\Bigr]$$
$$\prod_{j=1}^n\Bigl[ \int d{x^0}_j {{d}\over{d {x^0}_j}}
\Bigr]<0|T\Bigl( a_{{\bf p'}_m}({x^0}'_m),...,a_{{\bf p'}_1}({x^0}'_1)
a_{{\bf p}_n}^+({x^0}_n),...,a_{{\bf p}_1}^+({x^0}_1)\Bigr)|0>. \eqno(1.8)$$

The translation  (1.1a) for the 4D field in the momentum space
$\Phi'(q)=\Phi(q+h)$ generates the corresponding gauge 
transformation for $\Phi(x)$. 
In particular,   for the complex charged field
the four-momentum translation (1.1a)  produces the simplest 
gauge transformation

$${\Phi}'(x)=e^{ih_{\mu}x^{\mu}}\Phi(x); \ \ \ \ \ 
i{{\partial}\over{\partial{ x_{\mu}'} }}=i{{\partial}\over{\partial{ x_{\mu}} }}
+h_{\mu},\eqno(1.9)$$
where generally $h_{\mu}$ is a complex constant.

For the changeless real fields $\Phi(x)$
the gauge transformation (1.9)  is consistently defined 
for the pure imaginary $h_{\mu}=i r_{\mu}$ with the real $ r_{\mu}$. 
In particular, for the real scalar fields we get

$${\Phi}'(x)=e^{-r_{\mu}x^{\mu}}\Phi(x); \ \ \ \ \ 
{{\partial}\over{\partial{ x_{\mu}'} }}={{\partial}\over{\partial{ x_{\mu}} }}
+r_{\mu},\eqno(1.10)$$
The generalization of the  gauge transformations (1.10)
for the real scalar fields
within  the nonlinear $\sigma$ model was performed in \cite{Wei,Alf}.
These formulation we shall consider at end of Sect. 10.

The transformation of $\Phi(x)$ under the 
rotation and dilatation of $q_{\mu}$ (1.1b,c) can be reproduced
through the Fourier transformations in (1.6a) using
the inverse rotation and dilatation of $x_{\mu}$ in $exp(-iqx)$.

The doubling of the 5D fields $\varphi_{\pm}(x,x_5)$ 
in (I.4a,b) and (I.7) is generated by the intermediate 
projections onto domains of the 5D hyperboloids (I.3a,b).
These domains 
cover unambiguously the whole  area of $q_{\mu}$ 
and $q^2$ and cover the whole domain of the variables of the 
fields $\Phi_{\pm}(q)$ (I.10).  
Inversion $q_{\mu}'=-M^2 q_{\mu}/q^2$
transform the   domains of the variables  of $\varphi_1(q,q^2_5)$
into  the domain of the variables of $\varphi_1(q,q^2_5)$ and
vice versa. But inversion transforms also these fields, 
i.e. inversion  
replaces $\varphi_{1}(x,x_5)$ (1.4a) and  $\varphi_{2}(x,x_5)$ (I.4b)
with the $\varphi_{2}^{(I)}$  and  $\varphi_{1}^{(I)}$ 

$$\varphi_{1}{\stackrel{inversion}{\Longleftrightarrow}} \varphi^{(I)}_2
\ \ \ i.e.\ \ \ 
\varphi_{+}{\stackrel{inversion}{\Longleftrightarrow}} \varphi^{(I)}_+;\ \ \ 
\varphi_{-}{\stackrel{inversion}{\Longleftrightarrow}} -\varphi^{(I)}_-
,\eqno(1.12)$$.
where the upper index $ ^{(I)}$ denotes the inversion of the
corresponding operator. One needs to introduce this index because
the 4D equation of motion for the massive particles are not invariant
under the inversion. For instance, the equation of motion
$(q^2-m^2)\Phi(q)=j(q)$ after inversion transforms into new type equation 
$(M^2-q^2\ m^2/M^2)\phi^{(I)}(q)=q^2j^{(I)}(q)/M^2$.

\vspace{0.15cm}

\centerline{\bf{2.\  Domains of $q^2$ and $q^2_5$}}

\vspace{0.15cm}

\par

The invariant form $\kappa_A\kappa^A=0$ (I.2a)
of the $O(2,4)$ group can be represented for $q^2$
 as

$$q^2+M^2{{\kappa_{-}}\over{\kappa_{+}}}
=0,\eqno(2.1a)$$
where
$$q_{\mu}={{\kappa_{\mu}}\over{\kappa_+}};\ \ \ \
\kappa_{\pm}={{ \kappa_5\pm\kappa_6}\over {M}}. \eqno(2.1b)$$
It is convenient to use the auxiliary  fifth momentum $q_5^2$
instead of $\kappa_5/\kappa_6$  in (2.1a).
This procedure implies  a projection
of the 6D rotational invariant form
$\kappa_A\kappa^A=0$   into corresponding 5D forms.  
In order to cover unambiguously
 the whole domain $-\infty\le q^2\equiv q_{\mu}q^{\mu}
\le\infty$  we have  distributed  $q^2$ and 
corresponding  $q_5^2$ between the domains
of the two 5D hyperboloids

$$q^2+q_5^2=M^2 \ \ \ \ \ with \ \ \ \ \ \ \ \ \ \ \ \ \ \ \ \ \
 \ \ q_5^2 =M^2 { {2\kappa_{5}} \over {\kappa_{5}+\kappa_{6} } },\eqno(2.2a)$$

and

$$q^2-q_5^2=-M^2 \ \ \ \ \ with \ \ \ \ \ \ \ \ \ \ \ \ \ \ \ \ \ \
q_5^2 =M^2 { {2\kappa_{6}}\over {\kappa_{5}+\kappa_{6} } }.\eqno(2.2b)$$

The hyperboloids (2.2a,b) presents the 
simple intermediate 5D projection of the 6D cone 
(I.2a) into
the 4D momentum space with only one auxiliary variable $q_5^2$.

  $\kappa_+$ and $q^2_5$ in hyperboloids (2.2a,b) are
defined in positive. Consequently,
$\kappa_5$ on the hyperboloid (2.2a) and $\kappa_6$ 
in (2.2b)  are also positive. The domains of the variables 
 $q^2$,  $q^2_5$, $\kappa_5$ and $\kappa_6$ defined on the 6D cone (2.1a) and on
the corresponding 5D hyperboloids $q^2 \pm q_5^2=\pm M^2$ (2.2a,b)
are listed in Table 1. The border points $q^2=0$, $q^2=M^2$ and $q^2=-M^2$ are 
included in the domain ${\bf I}$ and ${\bf IV}$.

It must be emphasized, that the 
domain $M^2<q_5^2<2M^2$ of  the variables of $\phi(q,q_5^2)$ 
is excluded  by construction of $\varphi_{1,2}(x,x_5)$ (I.4a,b).
Correspondingly, the variables $\kappa_5$ and $\kappa_6$ cover also 
only the part of the 6D cone $\kappa_A\kappa^A=0$ (I.2a). 
The principal restriction for 
the auxiliary variables  $\kappa_5$, $\kappa_6$ and $q_5^2$ is that  
the corresponding $q^2$ must cover the whole domain $(-\infty, +\infty)$.
This property allows to construct unambiguously the 4D field
$\Phi_{\pm}(x)=\varphi_{\pm}(x,x_5=0)$.

\vspace{1.cm}

\begin{center}

{\bf Table 1}\ \ \ {\em Domains of $q^2$, $q_5^2$,
$\kappa_5$ and $\kappa_6$ placed  on the hyperboloids 
$q^2 \pm q_5^2=\pm M^2$ (2.2a,b) and on the surface (2.1a).}
\vspace{0.55cm}

\hspace{-0.75cm}
\begin{tabular}{|c|c|c|c|c|} \hline\hline
        &        {\bf I}            &         {\bf II}
            &      {\bf III}            &         {\bf IV }
\\    \hline
           & $q^2+q_5^2=M^2$ & $q^2-q_5^2=-M^2$ & $q^2+q_5^2=M^2$
& $q^2-q_5^2=-M^2$ \\ \hline
 $ q^2$     &   $0\le q^2═\le═M^2$   & $M^2< q^2< \infty$ &$-\infty<q^2<-M^2$
& $-M^2\le q^2< 0 $\\  \hline
 $ q_5^2$   &   $0\le q_5^2═\le M^2$ & $2M^2< q_5^2< \infty$ & $2M^2< q_5^2< \infty$ &  $0\le q_5^2< M^2$ \\ \hline
 $\kappa_5\&\kappa_6$ &  $0\le\kappa_5\le\kappa_6$   
& $\kappa_6> 0$;\ $\kappa_5<0$;\ $\kappa_5+\kappa_6>0$ & 
$\kappa_5> 0$;\  $\kappa_6<0$;\ $\kappa_5+\kappa_6>0$ &  
$0\le\kappa_6<\kappa_5$ \\ 
\hline \hline
\end{tabular}

\vspace{1.cm}

\end{center}

\par

The hyperboloids $q^2+q^2_5=M^2$ and $q^2-q^2_5=-M^2$ and the 
corresponding domains ${\bf I,III}$ and ${\bf II,IV}$ transforms also into 
each other by reflection $q^2\longleftrightarrow -q^2$ which is generated 
by transposition of the  variables $\kappa_5$ and $\kappa_6$

 $$ q^2\longleftrightarrow -q^2\ \ \ \ \ \ \ \ 
\kappa_5=\kappa_6,\ \ \ \kappa_6=\kappa_5;\ \ \ \ \
\kappa_+=\kappa_+,\ \ \ \kappa_-=-\kappa_-.\eqno(2.3)$$

The similar transpositions of 
the hyperboloids $q^2+q^2_5=M^2$ and $q^2-q^2_5=-M^2$ produces 
inversion $q_{\mu}^I=-M^2 q_{\mu}/q^2$ 
(1.1d) which is generated by transposition of 
the variables $\kappa_+$  and $\kappa_-$ according to (1.2d).

The choice of the 5D hyperboloids is not unique. 
For instance, instead of
the two 5D hyperboloids  (2.2a,b) one can take 
other hyperboloids 
$q^2\pm q_5^2=M^2$ with  
$q_5^2 = { {\pm 2 M^2\kappa_{5}} / {\kappa_{5}+\kappa_{6} } }$, where $\kappa_+$ 
is fixed  $\kappa_+={\cal M}/M$ and ${\cal M}$ is a mass  
parameter. But
 these domains of $q^2$ are not symmetric 
under the reflection and inversion of $q^2$
 unlike the domains in Table 1. Therefore
we do not consider them.


An other choice of the intermediate 5D projections 
presents the stereographic projection, where
 $\kappa_6=-1$,
and  the auxiliary  momenta 
$Q_{\mu}=q_{\mu}/(1-q^2)$ and $Q_4=(1+q^2)/(1-q^2)$, i.e.
$q_{\mu}=Q_{\mu}/(1+Q_4)$, $q^2=(Q_4-1)/(1+Q_4)$ are introduced
(see, for example,  eq. (13.43) in \cite{IZ}).
This choice of the variables require only one 
5D hyperboloid $Q^2-Q_4^2=-1$ with
 the 5 auxiliary variables $Q_{\mu}$ and $Q_4$
for the intermediate 5D projections. 
Certainly, one can represent this projection
via the considered  projections on the hyperboloids 
$q^2\pm q_5^2=\pm M^2$ (2.2a,b) where only one auxiliary 
variable $q^2_5$ is used.

\vspace{0.15cm}
\begin{center}

{\bf{3.\ 4D reductions of the 5D fields.}}
\end{center}
\vspace{0.15cm}

The  5D fields $\varphi_{\pm}=\varphi_{1}\pm 
\varphi_{2}$ (I.7) consist of the two parts
$\varphi_1$ (I.4a) and $\varphi_1$ (I.4b).
which satisfy the conditions (I.6),  
because their Fourier conjugates  
are defined on the 5D hyperboloids 
$q^2\pm q^2_5=\pm M^2$ (I.3a,b).
For $\varphi_{\pm}$ 
these conditions can be represented as

$${{\partial^2 \varphi_{\pm}(x,x_5)}\over{\partial x^{\mu}\partial x_{\mu}}}+
\Bigl({{\partial^2}\over{\partial x^5\partial x_5}}+M^2\Bigr)
\varphi_{\mp}(x,x_5)=0. \eqno(3.1)$$

Integration over $q_5^2$ in (I.4a,b) yields

$$\varphi_{\pm}(x,x_5)=\int {{d^4q}\over{(2\pi)^4}}e^{-iqx}
\Bigl[\phi(q,Q^2_1)\Lambda_1(q^2)e^{-iQ_1x_5}\pm
\phi(q,Q^2_2)\Lambda_2(q^2)e^{-iQ_2x_5}\Bigr].
\eqno(3.2)$$
The expression (3.2) can be represented
via the 5D convolution formula

$$\varphi_{\pm}(x,x_5)=\int d^5y \phi(x-y,x_5-y_5){\cal P}_{\pm}(y,y_5),
\eqno(3.3)$$
where $\phi(x,x_5)$ is the Fourier conjugate of 
the full 5D function $\phi(q,q_5^2)$ in (I.4a,b) and

$$ \Lambda_a(q^2) = \left\{ \begin{array}{ll}
\theta(q^2)\theta(M^2-q^2)+\theta(-q^2)\theta(-M^2-q^2)          
& \mbox{if $a=1$};\\
\theta(q^2)\theta(-M^2+q^2)+\theta(-q^2)\theta(M^2+q^2)         
& \mbox{if $a=2$},\end{array} \right. \eqno(3.4)$$

$$ Q_a^2 = \left\{ \begin{array}{ll}
M^2-q^2          
& \mbox{if $a=1$};\\
M^2+q^2         
& \mbox{if $a=2$},\end{array}\right.\ \ \ and\ \ \ Q_a=\sqrt{Q_a^2}. 
\eqno(3.5)$$

The operators 

$${\cal P}_{\pm}(x,x_5)={\cal P}_1(x,x_5)\pm {\cal P}_2(x,x_5)
\eqno(3.6a)$$

consist of the two  parts that are placed onto hyperboloids
$q^2+q_5^2=M^2$ and $q^2-q_5^2=-M^2$ 

$${\cal P}_a(x,x_5)=\left\{ \begin{array}{ll}
\int dq_5^2e^{-iq_5x_5} {{d^4q}/{(2\pi)^4} } e^{-iqx}\Lambda_1(q^2)
\delta(q^2+q_5^2-M^2)          
& \mbox{if $a=1$};\\
\int dq_5^2 e^{-iq_5x_5} {{d^4q}/ {(2\pi)^4} } e^{-iqx}
\Lambda_2(q^2)\delta(q^2-q_5^2+M^2)         
& \mbox{if $a=2$},\end{array} \right. \eqno(3.6b)$$

that satisfy the orthogonality 
and completeness conditions
at $x_5=0$ 

$$\int d^4y {\cal P}_{a}(x-y,0){\cal P}_{b}(y,0)
=\delta_{ab}{\cal P}_{a}(x,0);\eqno(3.7)$$ 

$${\cal P}_{+}(x,0)=\delta^{(4)}(x)\eqno(3.8)$$

The relation (3.3) presents the projections of the complete 
5D field $\phi(x,x_5)$
onto the two independent 5D fields $\varphi_{\pm}(x,x_5)$ which
for $x_5=0$ produce the 4D fields $\Phi_{\pm}(x)$ (I.8).
The Fourier conjugate $\Phi_{\pm}(q)$ (I.9) and the Fourier conjugate
of the 5D fields $\varphi_{\pm}$ (3.3)
consist of the same four parts that are given in (I.10) and (I.11).
Therefore from (3.3) we get

$$\Phi_{\pm}(x)=\varphi_{\pm}(x,x_5=0)
=\int {{d^4q}\over {(2\pi)^4}} e^{-iqx}\Bigl[
\phi(q,q_5^2=M^2-q^2)\Lambda_1(q^2)
\pm\phi(q,q_5^2=M^2+q^2)\Lambda_2(q^2)\Bigr],\eqno(3.9)$$

The equations (I.5) and (3.9) presents
the 4D reduction of the 6D field $\varsigma(\kappa,\kappa_-,\kappa_+)$  and 
 5D field $\phi(q,q^2_5)$ onto the 4D field $\Phi_{\pm}(q)$ (I.10).  

It must be noted that $\phi(q,M^2<q^2_5<2M^2)$ 
does not contribute  in 
the 4D fields $\Phi_{\pm}(x)$ because the region $M^2<q^2_5<2M^2$
is excluded from  the domains  in Table 1.

(3.3) and (3.8) allow to represent  (3.9)  m
as the 4D convolution formula

$$\varphi_{\pm}(x,x_5)=\int d^4y \Phi_{\pm}(x-y){\cal P}_{+}(y,x_5)
.\eqno(3.10)$$

that  determines the two 5D fields
$\varphi_{\pm}$ via the corresponding  4D fields $\Phi_{\pm}$.


\vspace{0.15cm}

\begin{center}

{\bf{4.\ 4D and   5D  equations of motion.} }

\end{center}
\vspace{0.15cm}
\par

According to the boundary condition (3.9) and the convolution
formula (3.10)
the 5D and 4D scalar fields $\varphi_{\pm}$ and $\Phi_{\pm}$ 
satisfy the similar equation of motion

$$\biggl({{\partial^2}\over{\partial x^{\mu}\partial x_{\mu}}}+m_{\pm}^2\biggr)
\varphi_{\pm}(x,x_5)=j_{\pm}(x,x_5),\eqno(4.1a)$$

$$\biggl({{\partial^2}\over{\partial x^{\mu}\partial x_{\mu}}}+m_{\pm}^2\biggr)
\Phi_{\pm}(x)=J_{\pm}(x)\eqno(4.1b)$$
  where  
$$J_{\pm}(x)=j_{\pm}(x,x_5=0).\eqno(4.2)$$

The 5D  source operators $j_{\pm}(x,x_5)$ in (4.1a) 
consist of the products of the 5D fields $\varphi_{\pm}$  
and their derivatives. But
the Fourier conjugate of  
the multiplication of $\varphi_{\pm}(x,x_5)$ are not 
placed on the hyperboloids $q^2\pm q_5^2=\pm M^2$. 
The source operators $j_{\pm}(x,x_5)$ in (4.1a)
must be placed on the hyperboloids $q^2\pm q_5^2=\pm M^2$
as well as $\varphi_{\pm}(x,x_5)$ in (3.1). Consequently,
 $j_{\pm}(x,x_5)$ must satisfy the additional 
5D constrains in analogy with (3.1) for 
$\varphi_{\pm}(x,x_5)$.
In order to obtain these conditions for $j_{\pm}$  we 
combine the equations (4.1a,b) and the conditions (3.1) as
$${{\partial^2 j_{\pm}(x,x_5)}\over{\partial x^{\mu}\partial x_{\mu}}}+
\Bigl({{\partial^2}\over{\partial x^5\partial x_5}}+M^2\Bigr)
j_{\mp}(x,x_5)=$$
$$\Bigl({ {\partial^2}\over { \partial x_{\mu} \partial x^{\mu}}}
{ {\partial^2}\over { \partial x_{5} \partial x^{5}}}-
{ {\partial^2}\over { \partial x_{5} \partial x^{5}}}
{ {\partial^2}\over { \partial x_{\mu} \partial x^{\mu}}}\Bigr)
\varphi_{\mp}(x,x_5)+(m_{\pm}^2-m_{\mp}^2)
{ {\partial^2 \varphi_{\pm}(x,x_5)}\over { \partial x_{\mu} \partial x^{\mu}}}
.\eqno(4.3)$$
For the independent  variables $x_{\mu}$ and $x_5$ 
the operator ${ {\partial^2}/ { \partial x_{\mu} \partial x^{\mu}}}$
and ${ {\partial^2}/ { \partial x_{5} \partial x^{5}}}$ commute.
Therefore we get
$${{\partial^2 { {\widetilde {j}}_{\pm}(x,x_5)}
\over{\partial x^{\mu}\partial x_{\mu}} }}+
\Bigl({{\partial^2}\over{\partial x^5\partial x_5}}+M^2\Bigr)
{\widetilde {j}}_{\mp}(x,x_5)=0,\eqno(4.4a)$$
where 
$$\widetilde {j}_{\pm}(x,x_5)=j_{\pm}(x,x_5)-m^2_{\pm}\varphi_{\pm}(x,x_5)
\eqno(4.4b)$$

Thus $\widetilde {j}_{\pm}$ satisfy also the same sourceless equation (3.1). 
Using (4.1a) one can rewrite (4.4a) as 

$${{\partial^2 
\over{\partial x^{\nu}\partial x_{\nu}} }}
{{\partial^2 { {\varphi}_{\pm}(x,x_5)}
\over{\partial x^{\mu}\partial x_{\mu}} }}+
{{\partial^2 
\over{\partial x^{\nu}\partial x_{\nu}} }}
\Bigl({{\partial^2}\over{\partial x^5\partial x_5}}+M^2\Bigr)
{\varphi}_{\mp}(x,x_5)=0,\eqno(4.4c)$$

which indicates that the Fourier conjugate of
${{\partial^2 { {\varphi}_{\pm}(x,x_5)}
/ {\partial x^{\mu}\partial x_{\mu}} }}$ and 
${\widetilde {j}}_{\pm}$
in  (4.1a) are placed on the hyperboloids $q^2\pm q_5^2=\pm M^2$.
Therefore, in analogy with (I.4a,b) we get

$$j_{1}(x,x_5)=\int {{d^4q}\over{(2\pi)^4}}dq_5^2e^{-iqx-iq^5x_5}
\delta(q^2+q_5^2-M^2)
\Bigl[\theta(q^2)\theta(M^2-q^2)+\theta(-q^2)\theta(-M^2-q^2)]
\Bigr]{\cal J}(q,q_5^2),\eqno(4.5a)$$

$$j_{2}(x,x_5)=
\int {{d^4q}\over{(2\pi)^4}}dq_5^2e^{-iqx-iq^5x_5}
\delta(q^2-q_5^2+M^2)
\Bigl[\theta(q^2)\theta(-M^2+q^2)+\theta(-q^2)\theta(M^2+q^2)\Bigr]
{\cal J}(q,q_5^2).\eqno(4.5b)$$
where ${\cal J}$ denote the full 5D source operator 
$${\cal J}(x,x_5)=\int {{d^5q}\over {(2\pi)}}e^{-iqx-iq_5x_5}{\cal J}(q.q_5^2)
\eqno(4.5c)$$

Afterwards the 5D sources $ {j}_{\pm}$ can be represented  as

$${ {j}}_{\pm}(x,x_5)=
\int {{d^4q e^{-iqx}}\over{(2\pi)^4}}\Bigl[
 {\cal J}(q,Q^2_1)\Lambda_1(q^2)e^{-iQ_1x_5}
\pm  {\cal J}(q,Q^2_2)\Lambda_2(q^2)e^{-iQ_2x_5}\Bigr],\eqno(4.6a)$$

that in analogy with (3.3) and (3.10) allow to construct the 5D sources 
${ {j}}_{\pm}(x,x_5)$ through the 4D sources $J_{\pm}(x)$ 
or the  full 5D sources 
${\cal J}$ using the following convolution formulas

$${ {j}}_{\pm}(x,x_5)=\int d^4y 
J_{\pm}(x-y){\cal P}_{+}(y,x_5),
\eqno(4.6b)$$ 

$$j_{\pm}(x,x_5)=
\int d^5y {\cal J}(x-y,x_5-y_5){\cal P}_{\pm}(y,y_5).\eqno(4.6c)$$

According to (4.6a) 
the sources $J_{\pm}(x)$ and
their Fourier conjugate    $J_{\pm}(q)$ 
consist of the four parts as well as  the 4D field $\Phi_{\pm}(q)$ in (I.10)

$$J_{\pm}(q)=\sum_{N=I,III} J_{N}(q)\pm \sum_{N=II,IV} J_{N}(q),\eqno(4.7)$$
where

$$J_{I}(q)= \theta(q^2)\theta(M^2-q^2){\cal J}(q,q^2_5=M^2-q^2);
\ \ \ \ \ J_{III}(q)
= \theta(-q^2)\theta(-M^2-q^2){\cal J}(q,q^2_5=M^2-q^2);\eqno(4.8a)$$ 
$$J_{II}(q)= \theta(q^2)\theta(-M^2+q^2){\cal J}(q,q^2_5=M^2+q^2);
\ \ \ \ \ J_{IV}(q)=
 \theta(-q^2)\theta(M^2+q^2){\cal J}(q,q^2_5=M^2+q^2)
,\eqno(4.8b)$$
where  the lower index  $N=I,II,III,IV$ in  ${ {J}_N}$ 
indicates the  domains of $q^2$ in Table 1.

In (4.6a) ${ {j}}_{\pm}(x,x_5)$ is constructed
through the 4D sources $J_{\pm}(q)$ (4.8a,b)
that  are defined  via the 4D fields $\Phi_{\pm}$,
i.e. (4.6a) allow to determine the 5D source $j_{\pm}$
through the 4D fields $\Phi_{\pm}$.
In (4.6b) ${ {j}}_{\pm}(x,x_5)$ is constructed is constructed via the 
5D sources ${\cal J}(x,x_5)$ (4.5c) which consists of the products of the complete
5D  fields $\phi(x,x_5)$.  
The 5D source  ${\cal J}$ (4.5c) can be determined via the 
fields $\phi$ according to the 5D extension of the Klein-Gordon equation


$$\biggl({{\partial^2}\over{\partial x^{\mu}\partial x_{\mu}}}+
   {\widetilde m}^2\biggr)\phi(x,x_5)={\cal J}(x,x_5).\eqno(4.9)$$

In the present formulation the terms with 
$\partial\varphi_{\pm}/\partial x_5$ are determined through the constrains which
ensure the consistency of the condition (3.1) and the equation of motion (4.1a).
These constrains are considered in the next section.
In (4.9) the projections of $\partial\phi_{\pm}/\partial x_5$
 can be included in ${\cal J}$. But these terms must be consistent with the
constrains for  $\partial\varphi/\partial x_5$.

The projections of the  5D field $\phi$ 
with mass ${\widetilde m}$ and  source ${\cal J}$
into two 5D fields $\varphi_{\pm}$ with the two different masses $m_{\pm}$
and the two sources $j_{\pm}$ are performed in (3.3) and
(4.6c) using the projection operators ${\cal P}_{\pm}$ (3.6a).
In  these projections  only the parts of
the complete 5D field and its source are used for construction of the 5D 
fields $\varphi_{\pm}$ and 
$j_{\pm}$  because the parts of these fields 
in the domains $N=I,II,III,IV$ in Table 1
cover only the part of the full 5D space.  

The input 5D fields $\phi$, ${\cal J}$ and
their projections $\varphi_{\pm}$, $j_{\pm}$
can be constructed within the various relativistic invariant
time models \cite{Fanchi,Land}.


It must be noted, that  $\partial \varphi_{\pm}(x,x_5)/\partial x_5$,
$j_{\pm}(x,x_5)$ and 
$m^2_{\pm}$ satisfy an additional conditions that can be obtained
combining (4.1a)  and (3.1) 
$$M^2\biggl[1+\Bigl({1\over M}{{\partial}\over{\partial x_5}}\Bigr)^2\biggr]
\varphi_{\pm}=m_{\mp}^2\varphi_{\mp}-j_{\mp}.\eqno(4.10)$$
The conditions (4.10) presents the relationship between
  $j_{\pm}$,  $m_{\pm}$ and
the operators  
$\biggl[1+\Bigl({1\over M}{{\partial}/{\partial x_5}}\Bigr)^2\biggr]
\varphi_{\pm}$. The factorization  of 
$\biggl[1+\Bigl({1\over M}{{\partial}/{\partial x_5}}\Bigr)^2\biggr]$
allows to determine  $\partial \varphi_{\pm}/\partial x_5 $ through
 $j_{\pm}(x,x_5)$, $m^2_{\pm}$ and vice versa. This problem will be considered 
in the next section.

\vspace{0.15cm}


\centerline{\bf{5.\  Constrains for  
${ {\partial \varphi_{\pm}}/ {\partial x^{5}}}$.}}
\vspace{0.15cm}

In order to obtain the constrains for  
${ {\partial \varphi_{\pm}}/ {\partial x^{5}}}$
we shall  factorize (4.10). 
The general form of the sought linear  conditions for
 ${ {\partial \varphi_{\pm}}/ { \partial x_{5} }}$
of the scalar charged fields are

$${i\over M}{{\partial \varphi_{\pm}}\over{\partial x_5}}=
\alpha_{\pm}\varphi_{+}+\beta_{\pm}\varphi_- +C_{\pm},\eqno(5.1)$$
where $C_{\pm}$ consist of the products of  $\varphi_{\pm}$.
The Fourier conjugate of these products 
are not located on the hyperboloids $q^2\pm q_5^2=\pm M^2$.
Therefore, $C_{\pm}$ are defined as

 $$C_{\pm}(x,x_5)=\int {\widetilde {\cal C}}_{\pm}(x-y,x_5-y_5)d^5y
{\cal P}_{+}(y,y_5). \eqno(5.2)$$

Substituting (5.1) in (4.10) one obtains

 $$m_+^2=-2M^2\alpha_+{{1-\alpha_+^2}\over {\beta_+}} 
\ \ \ \ \ 
j_+=M^2\biggl[{{1-\alpha_+^2}\over {\beta_+}} C_++\alpha_+C_-+{i\over M} 
{{\partial C_-}\over{\partial x_5} }\biggr];
\eqno(5.3a)$$
$$m^2_-=-2M^2\alpha_+\beta_+;\ \ \ \ \ j_-=M^2\biggl[\alpha_+C_++\beta_+C_-+
{i\over M} {{\partial C_+}\over{\partial x_5} }\biggr].\eqno(5.3b)$$
where are  using the conditions
$$\alpha_+^2+\alpha_-\beta_+=1;\ \ \ \ \ \ \ \ \ \ \ 
\beta_-^2+\alpha_-\beta_+=-1 \eqno(5.4)$$
that insures  reproduction of the mass terms in (4.10).

From (5.4) follows that $\alpha_+=\beta_-$
and  the condition  $m^2_{\pm}>0$ requires that $-1<\alpha_+<1$. 
If  $\alpha_+=-\beta_-$, then  $m_+=m_-=0$.

The relations (5.3a,b) determine $m^2_{\pm}$ through the three parameters
$M^2$, $\alpha_+$ and $\beta_+$. The same parameters, $C_{\pm}$ and 
 ${i/ M}\ {{\partial C_{\pm}}/{\partial x_5}}$
 determine the sources $j_{\pm}$.

For the neutral particles the analogue of (5.1) is 

$${1\over M}{{\partial \varphi_{\pm}}\over{\partial x_5}}=
\alpha_{\pm}\varphi_{+}+\beta_{\pm}\varphi_- +C_{\pm}
,\eqno(5.5)$$

Substituting this  condition in (4.10) we get

$$m_+^2=-2M^2\alpha_+{{1+\alpha_+^2}\over {\beta_+}} \ \ \ \ \ j_+=M^2\biggl[
{{1+\alpha_+^2}\over {\beta_+}}C_++\alpha_+C_-
+{1\over M} {{\partial C_-}\over{\partial x_5} }\biggr];
\eqno(5.6a)$$
$$m^2_-=2M^2\alpha_+\beta_+;\ \ \ \ \ j_-=M^2\biggl[\alpha_+C_++\beta_+C_-+
{1\over M} {{\partial C_+}\over{\partial x_5} }\biggr],\eqno(5.6b)$$
where the following conditions for the parameters are used

$$\alpha_+^2+\alpha_-\beta_+=-1;\ \ \ \ \ \ \ \ \ \ \ 
\beta_-^2+\alpha_-\beta_+=-1. \eqno(5.6c)$$ 

The ratio $m_+^2/m_-^2=-(\alpha_+^2+1)/\beta_+^2$ in this case is negative.
Therefore, the constrains (5.5) 
can not generate the positive $m_+^2$ and $m_-^2$ simultaneously
for the neutral fields.

The positive masses    $m_{\pm}^2\ge 0$ 
for the neuthral particles in (4.10) can be reproduced 
using the other kind of the constrains

$$\varphi_-=\alpha \varphi_++{\em  G}(\varphi_+,\varphi_-),\eqno(5.7a)$$
 where ${\em G}$ does not contain the linear over   
 $\varphi_{\pm}$ terms. For the sake of simplicity 
the dependence  over 
${\partial \varphi_{\pm}}/{\partial x_{\mu}}$ and 
${\partial \varphi_{\pm}}/{\partial x_{5}}$ in ${\em  G}$ (5.7a)
are omitted.

Acting  with
$\biggl[1+\Bigl({1\over M}{{\partial}/{\partial x_5}}\Bigr)^2\biggr]$
in (5.7a) and using the conditions

$$m_+^2=\alpha^2 m_-^2; \ \ \ \ \ 
{\em F}=
M^{-2}\biggl[1+\alpha {{m_-^2}\over {M^2}}+
\Bigl({1\over M}{{\partial}\over{\partial x_5}}\Bigr)^2\biggr]^{-1}
{\em  G}\eqno(5.7b)$$
one  determine $j_-$ via $j_+$ as 

$$\alpha j_--j_+={\em F}.\eqno(5.7c)$$

In particular, for the $\Phi_+^4$ model with
the 4D source
$J_+=a \Phi_+^2+b \Phi_+^3$ and mass $m_+$
the relations   (5.7a,b,c)  allow to reproduce
the  5D equations (4.1a) with the 5D source operator
$j_{\pm}$ and the masses $m_{\pm}^2$. These 5D equation generates 
the 4D equation of motion with 
$J_-=(J_++{\em F} )/\alpha$ and the real mass $m_-=m_+/\alpha$.

For the charged fields  it
is convenient to use the following  constrains

$$\biggl[1-{i\over M}{{\partial}\over{\partial x_5}}\biggr]
\varphi_{\pm}={\em C}_{\pm}.\eqno(5.8a)$$
where ${\em C}_{\pm}$ depends generally on the products of 
 $\varphi_{\pm}^+$ and $\varphi_{\pm}$.

$$M^2\biggl[1+{i\over M}{{\partial}\over{\partial x_5}}\biggr]
{\em C}_{\mp}=m_{\pm}^2\varphi_{\pm}-j_{\pm}.\eqno(5.8b)$$



Present formulation has in common with the other  
5D field-theoretical approaches within the relativistic invariant time 
method \cite{IZ,Fanchi,Land} with
 $x_5^2=x_0^2-{\bf x}^2\equiv x_{\mu}x^{\mu}$.
Unlike these 5D formulations our approach based on the 
invariance of the 6D and 5D forms (I.2a,b) and (I.3a,b) 
under the conformal transformations in the momentum space.
The main difference between the our conditions for
$\partial \varphi_{\pm}/\partial x_5$ (4.10), (5.1) and (5.7c)
and the  evaluation-type equations over the $x_5$ other authors
is that in the present formulation $\partial \varphi_{+}/\partial x_5$
is determined through the source  and mass of the coupling field
$j_{-}$ and $m_-$.
Nevertheless, one can use the 5D models in  \cite{IZ,Fanchi,Land}
for input 5D field $\phi(q,q^2_5)$
in (I.4a,b) and input 5D and 4D source operators 
$J_{\pm}(x)=j_{\pm}(x,x_5=0)$ in (4.1,b).


\vspace{0.15cm}

\centerline{\bf{6.\ The  5D and 4D Lagrangians.}}
\vspace{0.15cm}

In this section we shall consider  the 5D Lagrangians ${ {\cal L}}_{\pm}(x,x_5)$ 
for the two interacted scalar 5D fields $\varphi_{\pm}$
with the same quantum numbers, but with the different masses 
$m_{\pm}$ and sources $j_{\pm}(x,x_5)$. 
These Lagrangians must reproduce the 4D Lagrangians $L_{\pm}(x)$ 
at $x_5=0$ and the 5D equations of motion (4.1a).
The sought Lagrangian can be represented as  
$${ {\cal L}}_{\pm}(x,x_5)=
({ {\cal L}}_{\pm})_o(x,x_5)+
({{\cal L}}_{\pm})_{INT}(x,x_5)+
({{\cal L}}_{\pm})^C(x,x_5)\eqno(6.1)$$ 
where $({ {\cal L}}_{\pm})_o$ and $({ {\cal L}}_{\pm})_{INT}$
 denote the non-interacted and interacted parts of these  Lagrangians.
The third term $({{\cal L}}_{\pm})^C$  reproduce the constrains for the 
$\partial \varphi_{\pm}/\partial x_5$ 
in (5.1), (5.5), (5.7a) and (5.8a,b). Any one of these constrains 
together with the 5D equation of motion (4.1a)
generate the conditions (3.1). In particular,
$({{\cal L}}_{\pm})^C$  for the constrains  (5.8a,b) are

$${\cal L}_{\pm}^C=M^2|
{i\over M}{ {\partial \varphi_{\pm}}
\over { \partial x_5}}-\varphi_{\pm}+{\em C}_{\pm}|^2+
 M^2|{i\over M}{ {\partial {\em C}_{\pm}}\over { \partial x_5}}+
{\em C}_{\pm}
-{{ m_{\mp}^2\varphi_{\mp} -j_{\mp}}\over {M^2}}|^2,\eqno(6.2)$$
where $\delta {\cal L}_{\pm}^C/\delta {\em C}_{\pm}$ 
and $\delta {\cal L}_{\pm}^C/\delta[{{\partial {\em C}_{\pm}}
/ { \partial x_5}} ]$ reproduce
(5.8a) and (5.8b) correspondingly.


The 5D Lagrangians (6.1) can be simply constructed starting from  
the two 4D Lagrangians $L_{\pm}(x)$
for the two interacted scalar fields $\Phi_{\pm}(x)$

$$L_{\pm}(x)= (L_{\pm})_o(x)+({ L}_{\pm})_{INT}(x),\eqno(6.3)$$
where the non-interacting part $(L_{\pm})_o$

$$({ { L}}_{\pm})_o(x)=
{{\partial {\Phi_{\pm}}^{+}}\over{\partial x_{\mu}}}
{{\partial{\Phi_{\pm}}}\over{\partial x^{\mu}}}-m_{\pm}^2
{\Phi_{\pm}}^+{\Phi_{\pm}}\eqno(6.4a)$$

determines  the non-interacted part of the 5D Lagrangians (6.1)
$$({ {\cal L}}_{\pm})_o(x,x_5)=
{{\partial {\varphi_{\pm}}^{+}}\over{\partial x_{\mu}}}
{{\partial{\varphi_{\pm}}}\over{\partial x^{\mu}}}-m_{\pm}^2
{\varphi_{\pm}}^+{\varphi_{\pm}}.\eqno(6.4b)$$

The 5D source $j_{\pm}$ in (4.1a) can be constructed from the 4D source
$J_{\pm}$ (4.1b) using the convolution formula (4.6b). 
The sought 5D Lagrangian $({{\cal L}}_{\pm})_{INT}$
consists of the fields $\varphi_{\pm}$.
The replacement of the 4D fields $\Phi_{\pm}$ in 
 $(L_{\pm})_{INT}$ with the 5D fields $\varphi_{\pm}$ give

$$({{\cal L}}_{\pm})_{INT}(x,x_5)=\int d^4y ({\sf L}_{\pm})_{INT}(x-y,x_5)
{\cal P}_+(y,x_5),\ \ \ where\ \ \ 
({\sf L}_{\pm})_{INT}(x,x_5)=({ L}_{\pm})_{INT}
\Bigl(\Phi_{\pm}\Longleftrightarrow \varphi_{\pm}\Bigr).
\eqno(6.5)$$

The 5D sources  $ { j}_{\pm}$ satisfy the Euler-Lagrange equations

$$ { j}_{\pm}(x,x_5)={{\partial{ ({\cal L}_{\pm})_{INT}}}
\over{\partial {\varphi_{\pm}}^+} }
-{{d}\over{d x_{\mu}}} \biggl({ {\partial{ ({\cal L}_{\pm})_{INT}}}\over{\partial
{{[\partial {\varphi_{\pm}}^+} /{\partial x^{\mu}}]} } }\biggr),\eqno(6.6)$$


The other kind of the 5D sources $j_{\pm}$ can be obtained
from the general 5D Lagrangian
${\ell}(\phi,\phi^+)$ for the  5D scalar field $\phi(q,q_5^2)$ that parts
are used in (I.4a,b) and (3.2) for reproduction of  $\varphi_{\pm}$ 
 
$${\ell}(\phi,\phi^+)=
{{\partial {\phi}^{+}}\over{\partial x_{\mu}}}
{{\partial{\phi}}\over{\partial x^{\mu}}}+{\widetilde m}^2\phi^+\phi+
{\ell}_{INT}(\phi,\phi^+),\eqno(6.7)$$
where the terms with $\partial\phi /\partial x_5$ are included in ${\ell}_{INT}$.
The Lagrangian (6.7) generate the 5D equation of motion (4.9).
In order to reproduce $j_{\pm}$ from 
${\ell}(\phi,\phi^+)$ (6.7) one must determine  
$\partial\phi /\partial x_5$ according to constrains for 
$\partial\varphi_{\pm} /\partial x_5$ that are given
in $({{\cal L}}_{\pm})^C$ (6.1).
For this aim we shall consider the projection of the equation of motion (4.9)
according to (3.3) and (4.6c)

 $$\int d^5y
\Bigr[ {{\partial^2 \phi(x-y,x_5-y_5)}\over{\partial (x-y)^{\mu}\partial (x-y)_{\mu}}}
-{ {\cal J}}(x-y,x_5-y_5)+{\widetilde m}^2\phi(x-y,x_5-y_5)\Bigl]
{\cal P}_{\pm}(y,y_5)=0\eqno(6.8a)$$
This equation has the form of the equation (4.1a) if 
$$\int d^5y { {\cal J}}(x-y,x_5-y_5){\cal P}_{\pm}(y,y_5)
=j_{\pm}(x,x_5)-
\Bigl(m^2_{\pm}-{\widetilde m}^2\Bigr)\varphi_{\pm}(x,x_5)\eqno(6.8b)$$

The sources $j_{\pm}$ in (6.8a,b) 
are defined in (4.6c) via the projections of 
 ${\cal J}$. The complete 5D source  ${\cal J}$
  consists of the products of the fields $\phi$ and their derivatives. 
Therefore,  ${\cal J}$ 
can not be reduced to the combination of the fields $\varphi_{\pm}$
that are the parts of $\phi(x,x_5)$. 
Thus in opposite to  $j_{\pm}$ constructed from the Lagrangian
(6.5), the projections of ${\cal J}$ in (6.8a,b) does not satisfy the equation 
(6.6) and the Lagrangian  . 
${\ell}(\phi,\phi^+)$ (6.7) can not be reduced to the Lagrangians (6.1)
$({\cal L}_{\pm}$.
The consistency conditions 
of the projections of ${\cal J}$ in (6.8a,b)
in the equation of motion (4.1a) and the condition (3.1) can be obtained 
using the constrains
for $\partial\varphi /\partial x_5$ considered in the previous section.


The third form  of the 5D Lagrangians (6.1) reproduce exactly
the constrains
(3.1) and (5.8a,b) with the a priory given source $j_{\pm}$

$${\cal L}(x,x_5)=
{{\partial {\varphi_{+}}^{+}}\over{\partial x_{\mu}}}
{{\partial{\varphi_{+}}}\over{\partial x^{\mu}}}+M^2
{\varphi_{-}}^+{\varphi_{+}}+({{\cal L}}_{+})^C+
{{\partial {\varphi_{-}}^{+}}\over{\partial x_{\mu}}}
{{\partial{\varphi_{-}}}\over{\partial x^{\mu}}}+M^2
{\varphi_{+}}^+{\varphi_{-}}+({{\cal L}}_{-})^C
.\eqno(6.11)$$
where 
the Lagrangians $({{\cal L}}_{\pm})^C$ generate the 
conditions (5.8a,b) and variation over 
${\varphi_{\pm}}^{+}$ and 
${{\partial {\varphi_{\pm}}^{+}}/{\partial x_{\mu}}}$
produces  (3.1).
The combination of (3.1) and (5.8a.b) gives 
the equation of motion (4.1a).

 

\vspace{0.15cm}

\begin{center}

{\bf{7.\ The 4D and 5D equation
of motion for the fermion fields with 
the electromagnetic interactions.}}

\end{center}

\vspace{0.15cm}

The 4D and 5D equation of motion for the two  fermion fields 
$\Psi_+(x)=\psi_+(x,x_5=0)$ and $\Psi_-(x)=\psi_-(x,x_5=0)$ 
with the same quantum numbers but with the different masses 
$m_{\pm}$ and source operators can be represented in 
 analogy with the scalar fields. In order to place
the Fourier conjugate of the
5D Dirac fields  $\psi_{1,2}(x,x_5)=1/2\Bigl(\psi_{\pm}(x,x_5)
\pm \psi_{\pm}(x,x_5)\Bigr)$ on the hyperboloids $q^2\pm q_5^2=\pm M^2$ the 
fields $\psi_{+}(x,x_5)$ and   $\psi_-(x,x_5)$ must satisfy  
the condition (3.1)  
$$
{{\partial^2 \psi_{\pm}(x,x_5)}\over{\partial x^{\mu}\partial x_{\mu}}}+
 \Bigl({{\partial^2}\over{\partial x^5\partial
x_5}}+M^2\Bigr)\psi_{\mp}(x,x_5)=0. \eqno(7.1)$$ 
Similarly with  $\varphi_{\pm}$ in (I.10),
we divide  the Fourier conjugate of $\psi_{\pm}(x,x_5)$ into four parts 
defined in the domains I,II,III 
and IV  in Table 1   
$$\psi_{I}(q,q_5^2)=\theta(q^2)\theta(M^2-q^2)\theta(M^2-q_5^2)\Upsilon(q,q_5^2);
\ \ \ \psi_{II}(q,q_5^2)= \theta(q^2)\theta(-M^2+q^2)\theta(q_5^2-2M^2)
\Upsilon(q,q^2_5);$$
$$\psi_{III}(q,q_5^2)= \theta(-q^2)\theta(-M^2-q^2)\theta(q_5^2-2M^2)
\Upsilon(q,q^2_5);\ \ \
\psi_{IV}(q,q_5^2)= \theta(-q^2)\theta(M^2+q^2)\theta(M^2-q_5^2)
\Upsilon(q,q_5^2),\eqno(7.2)$$
Then as in (3.2) for $\varphi_{\pm}(x,x_5)$, 
for $\psi_{\pm}(x,x_5)$ we have 

$$\psi_{\pm}(x,x_5)
=\int {{d^4q}\over {(2\pi)^4}} e^{-iqx}\biggl[\sum_{N=I,III}
\psi_{N}(q,Q_1^2)e^{-iQ_1x_5}\pm \sum_{N=II,IV}\psi_{N}(q,Q_2^2)e^{-iQ_2x_5}
\biggr],\eqno(7.3)$$
and $\psi_{\pm}(x,x_5)$ satisfy
automatically  the condition (7.1).

For $x_5=0$  $\psi_{\pm}$  (7.3) generate the physical 4D fields 
$\Psi_{\pm}(x)$

$$\Psi_{\pm}(x)=\psi_{\pm}(x,x_5=0)=
\int {{d^4q}\over {(2\pi)^4}} e^{-iqx}\biggl[
\sum_{N=I,III}\Psi_{N}(q)\pm \sum_{N=II,IV}\Psi_{N}(q)\biggr],
\eqno(7.4)$$

where 

$$\Psi_{I}(q)= \psi_I(q,q^2_5=M^2-q^2);\ \ \
\Psi_{II}(q)= \psi_{II}(q,q^2_5=M^2+q^2);$$

$$\Psi_{III}(q)= \psi_{III}(q,q^2_5=M^2-q^2);\ \ \
\Psi_{IV}(q)= \psi_{IV}(q,q^2_5=M^2+q^2)\eqno(7.5a)$$

and

$$\Psi_{\pm}(q)=\sum_{N=I,III}\Psi_{N}(q)\pm \sum_{N=II,IV}\Psi_{N}(q).
\eqno(7.5b)$$

According to (7.4) and (7.5a,b) 
the Fourier conjugate of the 5D fields $\psi_{\pm}(x,x_5)$ (7.3) 
contains  the same four parts  as the  Fourier conjugate of the 4D fields
$\Psi_{\pm}(x)$. Therefore, one can represent the
$\psi_{\pm}(x,x_5)$ through $\Psi(x)$ as

$$\psi_{\pm}(x,x_5)=\int d^4y\Psi_{\pm}(x-y)
{\cal P}_+(y,x_5).\eqno(7.6)$$

The convolution formula (7.6) is similar to (3.10) for 
the scalar fields $\varphi_{\pm}(x,x_5)$.

The equation of motion for the 5D fields $\psi_{\pm}$ can be derived using
the gauge transformation of the 5D Dirac equations
for the two non-interacting fields $\psi_{\pm}^{(o)}$ with the 
different masses $m_{\pm}$

$$\Bigl(i\gamma_{\mu}{{\partial }\over{\partial x_{\mu}}}
-m_{\pm}\biggr)\psi_{\pm}^{(o)}(x,x_5)=0,\eqno(7.7)$$

where

$$
\psi_{\pm}^{(o)}(x,x_5)=\int d^4 y  \exp{\Bigl(ie\Lambda(x-y,x_5)\Bigr)}
\psi_{\pm}(x-y,x_5){\cal P}_{+}(y,x_5);\eqno(7.8a)$$ 
$$ie{ A}^{\mu}(x,x_5)=
\exp{\Bigl(-ie\Lambda(x,x_5)\Bigr)}{{\partial}\over{\partial x_{\mu}}}
\exp{\Bigl(ie\Lambda(x,x_5)\Bigr)}.\eqno(7.8b)$$  

Substituting (7.8a,b) into (7.7) we obtain

$$\int d^4y e^{ie\Lambda(x-y,x_5)}\biggl(
i\gamma_{\mu}{{\partial }\over{\partial x_{\mu}}
}-e\gamma_{\mu}{ A}^{\mu}(x-y,x_5)
-m_{\pm}\biggr)\psi_{\pm}(x-y,x_5)
{\cal P}_{+}(y,x_5)=0.\eqno(7.9)$$

According to (3.8) the projection operator
 ${\cal P}_{+}(y,x_5)$  for $x_5=0$ transforms into $\delta^4(y)$. 
Therefore, for $x_5=0$ the 5D non-local equations (7.9) 
generate the usual  4D Dirac equation for the fermion field with the 
electromagnetic interaction

$$
\biggl(i\gamma_{\mu}{{\partial }\over{\partial x_{\mu}}}-e\gamma_{\mu}
{\sf A}^{\mu}(x)-m_{\pm}\biggr)\Psi_{\pm}(x)=0.\eqno(7.10)$$
where  ${\sf A}^{\mu}(x)={ A}^{\mu}(x,x_5=0)$,
 $\Psi_{\pm}^{(o)}(x)=\psi_{\pm}^{(o)}(x,x_5=0)$  are the asymptotic
$''in''$ or $''out''$ fields which satisfy the equations

$$\Bigl(i\gamma_{\mu}{{\partial }\over{\partial x_{\mu}}}
-m_{\pm}\biggr)\Psi_{\pm}^{(o)}(x)=0;\eqno(7.11a)$$

$$\Psi_{\pm}^{(o)}(x)=\exp{\Bigl(ie\Lambda(x,0)\Bigr)}\Psi_{\pm}(x);\ \ \ \ \ \ 
ie{\sf A}^{\mu}(x)=
\exp{\Bigl(-ie\Lambda(x,0)\Bigr)}{{\partial}\over{\partial x_{\mu}}}
\exp{\Bigl(ie\Lambda(x,0)\Bigr)}.\eqno(7.11b)$$

The Fourier conjugate of the 5D gauge fields 
${ A}^{\mu}(x,x_5)$ are embedded on  the hyperboloids $q^2\pm q_5^2=\pm M^2$
and they satisfy the condition (3.1)
 As well as the scalar fields $\varphi_{\pm}$. Due to gauge invariance
one can use any combination of  ${ A}^{\mu}_+$ and
${ A}^{\mu}_-$ in the gauge transformations (7.8b). 
Therefore, in (7.8a,b)-(7.11a,b) and in the following formulas 
the lower indexes  ${\pm}$ 
of $\Lambda$, ${\sf A}^{\mu}$ and ${ A}^{\mu}$ are omitted.

The 5D representations 
of the interacted and non-interacted 5D fields
$\psi_{\pm}$ and $\psi_{\pm}^{(o)}$ are single-valued
determined through the their 4D reductions  $\Psi_{\pm}$ and $\Psi_{\pm}^{(o)}$
according to (7.6).
Therefore 
$\psi_{\pm}^{(o)}$ satisfies also the condition (7.1)
$$
{{\partial^2 \psi_{\pm}^{(o)}(x,x_5)}\over{\partial x^{\mu}\partial x_{\mu}}}+
 \Bigl({{\partial^2}\over{\partial x^5\partial
x_5}}+M^2\Bigr)\psi_{\mp}^{(o)}(x,x_5)=0. \eqno(7.12)$$
 
 The gauge transformation (7.8a) in (7.12) together with (7.1)
 yields 

$$\eta_{\pm}(x,x_5)=-\eta_{\mp}^{(5)}(x,x_5),\eqno(7.13)$$
where 

$$\eta_{\pm}(x,x_5)=\biggl(
ie{{\partial }\over{\partial x^{\mu}}}{ A}^{\mu}(x,x_5)
+ie{ A}^{\mu}(x,x_5){{\partial }\over{\partial x^{\mu}}}
-e^2{ A}^{\mu}(x,x_5){ A}_{\mu}(x,x_5)\biggr)\psi_{\pm}(x,x_5),
\eqno(7.14a)$$

and the auxiliary source operator $\eta_{\pm}^{(5)}$
 is defined via ${ A}^{5}$ as

$$\eta_{\pm}^{(5)}(x,x_5)=\biggl(
ie{{\partial }\over{\partial x^{5}}}{ A}^{5}(x,x_5)
+ie{ A}^{5}(x,x_5){{\partial }\over{\partial x^{5}}}
-e^2{ A}^{5}(x,x_5){ A}_{5}(x,x_5)\biggr)\psi_{\pm}(x,x_5),
\eqno(7.14b)$$

$$ie{ A}^{5}(x,x_5)=
\exp{\Bigl(-ie\Lambda(x,x_5)\Bigr)}{{\partial}\over{\partial x_{5}}}
\exp{\Bigl(ie\Lambda(x,x_5)\Bigr)}.\eqno(7.14c)$$  

Condition (7.13) allows to define ${ A}^{5}$ through $\eta_{\pm}$ and $\psi_{\pm}$.



The complete 5D fields $\Upsilon(x,x_5)$ 
and their Fourier conjugate $\Upsilon(q,q_5^2)$ in (7.2) can be defined
in the relativistic invariant time models \cite{Fanchi,Land}.
The projections of $\Upsilon(x,x_5)$ and $\Upsilon^{(o)}(x,x_5)$
 onto hyperboloids $q^2\pm q_5^2=\pm M^2$
can be determined also via the 5D convolution formulas
  
$$\psi_{\pm}(x,x_5)=\int d^5y { \Upsilon}(x-y,x_5-y_5)
{\cal P}_{\pm}(y,y_5);\ \ \ 
\psi_{\pm}^{(o)}(x,x_5)=\int d^5y { \Upsilon}^{(o)}(x-y,x_5-y_5)
{\cal P}_{\pm}(y,y_5),\eqno(7.15)$$

\vspace{0.15cm}


\begin{center}

{\bf{8.\ Constrains for $\partial\psi_{\pm}/\partial x_5$.
}} 

\end{center}

\vspace{0.15cm}


The consistency condition between the  
 5D Dirac equations (7.9) and the constrains
(7.1) can be represent through the linear equations for
 $\partial\psi_{\pm} /\partial x_5$. For this aim it is 
convenient to rewrite the  (7.9) 
in the form of the  Klein-Gordon equations.
Action of the operator 
$i\gamma_{\mu}{{\partial }/{\partial x_{\mu}}}+m_{\pm}$
 on (7.7) produces the following Klein-Gordon equations

$$\biggl({{\partial^2}\over{\partial x^{\mu}\partial x_{\mu}}}+m_{\pm}^2\biggr)
\psi_{\pm}^{(o)}(x,x_5)=0.\eqno(8.1)$$

Then the gauge transformation (7.8a) yields

$$\int d^4y e^{ie\Lambda(x-y,x_5)}\biggl[\Bigl(
{{\partial^2 }\over{\partial x_{\mu} \partial x^{\mu}}}
-m_{\pm}^2\Bigr)\psi_{\pm}(x-y,x_5)-\eta_{\pm}(x-y,x_5)\biggr]
{\cal P}_{+}(y,x_5)=0,\eqno(8.2)$$
where $\eta_{\pm}(x,x_5)$ is defined in (7.14a).

One can replace (7.1) with
 
$$\int d^4y e^{ie\Lambda(x-y,x_5)}\biggl[
{{\partial^2 \psi_{\pm}(x-y,x_5)}\over{\partial x_{\mu} \partial x^{\mu}}}
+\Bigl(M^2+ {{\partial^2 }\over{\partial x_{5} \partial x^{5}}}\Bigr)
\psi_{\mp}(x-y,x_5)\biggr]
{\cal P}_{+}(y,x_5)=0.\eqno(8.3)$$
 
 Afterwards  we obtain the following consistency 
condition of (8.3) and (8.2) for the 5D Dirac fields 

$$\int d^4y e^{ie\Lambda(x-y,x_5)}\Biggl\{ \biggl[
M^2+\Bigl( { {\partial}\over{\partial x_5}} \Bigr)^2\biggr]
\psi_{\pm}(x-y,x_5)-m_{\mp}^2\psi_{\mp}(x-y,x_5)+
\eta_{\mp}(x-y,x_5)\Biggr\}{\cal P}_{+}(y,x_5)=0.\eqno(8.4)$$

According to  (8.4)  
$\Bigr[M^2+\bigl( \partial /{\partial x_5}\bigr)^2\Bigr]\psi_{+}$
is  determined by the source  and mass of the 
$\psi_{-}$ fields in (8.2)
and vice versa. Unlike  (4.10) the condition (8.4)
is given in the form of the 4D convolution formula.


One can factorise (8.4) in the same way as  (4.10) using
the constrain for $\partial\ \psi_{\pm}/\partial x_5$.
In particular, 

$${i\over M}{{\partial \over{\partial x_5}}\psi_{\pm}}
=\alpha_{\pm}\psi_{+}+\beta_{\pm}\psi_- +C_{\pm}
,\eqno(8.5)$$
where $C_{\pm}$ are defined onto hyperboloids $q^2\pm q_5^2=\pm M^2$
according to (5.2) 
and $C_{\pm}$ does not contain the linear terms over $\psi_{\pm}$.

The  constrains   (8.5) allow to construct the relationship 
between the masses  $m_{\pm}^2$, sources  $j_{\pm}$
and the parameters 
$M^2$, $\alpha_+$, $\beta_+$ and $C_{\pm}$ in the same way as in  (5.3a,b).
In particular, 
the choice of the parameters according to (5.4) gives

 $${ {m_{+}^2}\over {m_{-}^2 }}={ {1-\alpha_+^2}\over {\beta_+^2}};\ \ \ \ \
 { {m_{+}^2 m_{-}^2}\over {4M^2} }=\alpha_+^2(1-\alpha_+^2);\ \ \ \ \
 \alpha_+^2={1\over 2}\pm {1\over 2}
\sqrt{1- {{ m_+^2 m_-^2}\over{M^4} }  }\eqno(8.6)$$

The ratio $m_{+}^2/m_-^2$ and the product $m_{+}^2m_{+}^2/4M^4$
in (8.6) are positive if 
$$0<\alpha_+^2<1; \ \ \ \ \ \ \ \ \ \ \ \ \ \ \ \ \ \ 
  { {m_{+}^2 m_{+}^2}\over {M^4} }\le 1.\eqno(8.7)$$
Therefore, the   constrain (8.5)   
 generate the positive masses $m_{+}^2$ and $m_-^2$ simultaneously 
if $M^2\ge m_{+}m_-$.

\vspace{0.15cm}


\begin{center}

{\bf{9.\ 5D extension of the standard $SU(2)\times U(1)$ model for the electron and muon. 
 }} 

\end{center}

\vspace{0.15cm}


As example of  unification of  the 4D fields with the same quantum numbers 
and different masses and sources, we shall consider
the relationship between the 4D electron and muon fields
$\Psi_+\equiv \Psi_{el}$ and   $\Psi_-\equiv \Psi_{muon}$, $\Psi_+\equiv \Psi_{el}$ and   
$\Psi_-\equiv \Psi_{muon}$ according to (7.3)-(7.6). 
In the standard 
Weinberg-Salam $SU(2)\times U(1)$ model \cite{Weinberg} we have
the following 4D equations of motion

$$\Bigl(i\gamma_{\mu}{{\partial }\over{\partial x_{\mu}}}-m_{el}\Bigr)\Psi_{el}=
{\cal J}_{el};\ \ \ \ \ \ \ \ \ \ \ \ \ 
\Bigl(i\gamma_{\mu}{{\partial }\over{\partial x_{\mu}}}-m_{muon}\Bigr)
\Psi_{muon}={\cal J}_{muon};\eqno(9.1a)$$
$${\cal J}_{el}=\Bigl(-e\gamma_{\mu}{\sf A}^{\mu}
+{{g^2-{g'}^2}\over{2\sqrt{g^2+{g'}^2} }}\gamma_{\mu}{\sf Z}^{\mu}
{{1+\gamma_5}\over 2}
+g'\gamma_{\mu}{\sf Z}^{\mu}{{1-\gamma_5}\over 2}\Bigr)\Psi_{el}
+{g\over {\sqrt{2} }}\gamma_{\mu}{\sf W}^{\mu}{{1+\gamma_5}\over 2}\nu_{el}
,\eqno(9.1b)$$ 
$${\cal J}_{muon}=\Bigl(-e\gamma_{\mu}{\sf A}^{\mu}
+{{g^2-{g'}^2}\over{2\sqrt{g^2+{g'}^2} }}\gamma_{\mu}{\sf Z}^{\mu}
{{1+\gamma_5}\over 2}
+g'\gamma_{\mu}{\sf Z}^{\mu}{{1-\gamma_5}\over 2}\Bigr)\Psi_{muon}
+{g\over {\sqrt{2} }}\gamma_{\mu}{\sf W}^{\mu}{{1+\gamma_5}\over 2}\nu_{muon}
,\eqno(9.1c)$$
where $\nu_{el}$ and  $\nu_{muon}$ denote the corresponding neutrino fields, 
${\sf W}^{\mu}$ and ${\sf Z}^{\mu}$
stands for the charged and neutral vector fields, 
${\sf A}^{\mu}$  is the photon field 

$$g=-{e\over{ sin\theta}};\ \ \ \ \ g'=-{e\over{ cos\theta}};\ \ \ \ \  
sin^2\theta=0.222\pm 0.011\eqno(9.2)$$

For derivation of (9.1a,b,c) was used the 4D  
gauge transformation of the 
neutrino-electron and neutrino-muon doublets  according to 
 (21.3.12)-(21.3.20) in \cite{Weinberg}.

$$\sum_{\alpha}{\cal A}^{\mu}_{\alpha}(x){\em t}_{\alpha}+y { B}^{\mu}=
\exp{\Bigl(\Lambda(x)\Bigr)}{{\partial}\over{\partial x_{\mu}}}
\exp{\Bigl(\Lambda(x)\Bigr)}.\eqno(9.1d)$$ 

$$
{\sf W}^{\mu}={1\over{\sqrt{2}}}
\Bigl({\cal A}_1^{\mu}+i{\cal A}_2^{\mu}\Bigr);
\ \ \ 
{\sf Z}^{\mu}=cos\theta\ {\cal A}_3^{\mu}+sin\theta\ { B}^{\mu};
\ \ \ 
{\sf A}^{\mu}=-sin\theta\ {\cal A}_3^{\mu}+cos\theta\ { B}^{\mu};
\eqno(9.1e)$$

The direct 5D extension of the 4D  Dirac equations (9.1a) can be
obtained in the same way as (7.9)
using the 4D gauge transformations (7.8a,b) with the common parameter $x_5$

$$\int d^4y e^{ie\Lambda(x-y,x_5)}\biggl[\Bigl(
i\gamma_{\mu}{{\partial }\over{\partial x_{\mu}}}
-m_{el}\Bigr)\psi_{el}(x-y,x_5)-{\widetilde \eta}_{el}(x-y,x_5)
\biggr]{\cal P}_{+}(y,x_5)=0.\eqno(9.3a)$$

$$\int d^4y e^{ie\Lambda(x-y,x_5)}\biggl[\Bigl(
i\gamma_{\mu}{{\partial }\over{\partial x_{\mu}}
}-m_{muon}\Bigr)\psi_{muon}(x-y,x_5)-{\widetilde \eta}_{muon}(x-y,x_5)
\biggr]{\cal P}_{+}(y,x_5)=0,\eqno(9.3b)$$

where $\Lambda(x,x_5=0)=\Lambda(x)$
For $x_5=0$  the 5D fields and sources in (9.3a,b) are
reduced into the 4D  fields and sources  

$$\Psi_{el,}(x)=\psi_{el}(x,x_5=0);\ \ \ \ \ \ \ \ \ \ 
\Psi_{muon}(x)=\psi_{muon}(x,x_5=0);\eqno(9.3c)$$
$${\widetilde \eta}_{el}(x,x_5=0)={\cal J}_{el}(x);\ \ \ \ \ \ \ \ \ \ 
{\widetilde \eta}_{muon}(x,x_5=0)={\cal J}_{muon}(x),\eqno(9.3d)$$  

where
 
$${\widetilde \eta}_{el}(x,x_5)=\Bigl(-e\gamma_{\mu}{ A}^{\mu}
+{{g^2-{g'}^2}\over{2\sqrt{g^2+{g'}^2} }}\gamma_{\mu}{ Z}^{\mu}
{{1+\gamma_5}\over 2}
+g'\gamma_{\mu}{ Z}^{\mu}{{1-\gamma_5}\over 2}\Bigr)\psi_{el}
+{g\over {\sqrt{2} }}\gamma_{\mu}{ W}^{\mu}{{1+\gamma_5}\over 2}\nu_{el}
,\eqno(9.4a)$$ 
$${\widetilde \eta}_{muon}(x,x_5)=\Bigl(-e\gamma_{\mu}{ A}^{\mu}
+{{g^2-{g'}^2}\over{2\sqrt{g^2+{g'}^2} }}\gamma_{\mu}{ Z}^{\mu}
{{1+\gamma_5}\over 2}
+g'\gamma_{\mu}{ Z}^{\mu}{{1-\gamma_5}\over 2}\Bigr)\psi_{muon}
+{g\over {\sqrt{2} }}\gamma_{\mu}{ W}^{\mu}
{{1+\gamma_5}\over 2}\nu_{muon}
,\eqno(9.4b)$$

It is easy to check, that for $x_5=0$ equation (9.3a,b) transforms into
(9.1a) due to ${\cal P}_+(x,x_5=0)=\delta(x)$. 

In order to place the  Fourier conjugate of the
5D  sources (9.4a,b) onto hyperboloids  $q^2\pm q_5^2=\pm M^2$ 
we shall use the following convolution formula   

$$\eta_{el}(x,x_5)=\int d^4y{\widetilde \eta}_{el}(x-y,x_5)
{\cal P}_+(y,x_5);\ \ \ 
\eta_{muon}(x,x_5)=\int d^4y{\widetilde \eta}_{muon}(x-y,x_5)
{\cal P}_+(y,x_5).\eqno(9.5a)$$

The exact form of 
${\cal P}_+$ (3.6a,b) allows to rewrite (9.5a) as 

$$\eta_{el}(x,x_5)=\int d^4y{\cal J}_{el}(x-y)
{\cal P}_+(y,x_5);\ \ \ 
\eta_{muon}(x,x_5)=\int d^4y{\cal J}_{muon}(x-y)
{\cal P}_+(y,x_5)
.\eqno(9.5b)$$

 $\psi_{el}$ and $\psi_{muon}$ satisfy the condition (7.1).
Therefore, there are two different ways for construction of the 5D 
fields $\psi_{el}$ and $\psi_{muon}$



{\sf 1.} If the electron and muon fields are completely independent, then 
the constrain   (7.1) can be reproduced through 
doubling of the electron and muon fields separately, 
i.e. through $(\psi_{el})_{\pm}$ and $(\psi_{muon})_{\pm}$,
where $(\psi_{el})_-$ and $(\psi_{muon})_-$ corresponds to the ``electron''
and ``muon`` with the  negative  or imaginary mass. 
This doubling of the electron 
and muon states  can be realized via the  appropriate constrains (8.5)
and the equations of motion  (7.9).


{\sf 2.} 
If the 5D fields $\psi_{el}$ and $\psi_{muon}$  consists of 
the same parts,
then in (7.1) and (7.9) $\psi_{el}\equiv \psi_+$ and
$\psi_{muon}\equiv \psi_-$ 
and for their 4D reductions  we have

$$\Psi_{el}(x)=
\int {{d^4q}\over {(2\pi)^4}} e^{-iqx}\biggl[
\sum_{N=I,III}\Psi_{N}(q)+ \sum_{N=II,IV}\Psi_{N}(q)\biggr],
\eqno(9.6a)$$
$$\Psi_{muon}(x)=
\int {{d^4q}\over {(2\pi)^4}} e^{-iqx}\biggl[
\sum_{N=I,III}\Psi_{N}(q)- \sum_{N=II,IV}\Psi_{N}(q)\biggr].
\eqno(9.6b)$$


The equations (9.6a,b) unify
 the 4D electron and muon Heisenberg fields
which satisfies the 4D equation of motion (9.1a).
 Despite of the mixing of the $\Psi_{el}(x)$ and $\Psi_{muon}(x)$
in (9.6a,b) the 4D equations of motion   (9.1a)
are the same as in  the standard  model \cite{Weinberg},
where  the electron-muon coupling is strongly suppressed.
Therefore, the perturbation series constructed in the framework of the 
Weinberg-Salam $SU(2)\times U(1)$ theory and the perturbation series based 
on the equation of motion (9.1a,b,c) with the mixed fields (9.6a,b) 
coincides.






The common structure of the interacted Heisenberg fields
$\Psi_{el}(x)$ and $\Psi_{muon}(x)$ in (9.6a,b)  
is important for the theories beyond Weinberg-Salam model. 
In particular, within  the non-perturbative formulation
the  functional integrals  with the  4D  
electron and muon fields 
$\int {\cal D}(\Psi_{el}){\cal D}({\overline \Psi}_{el})\Bigl[...\Bigr]$ 
and $\int {\cal D}(\Psi_{muon}){\cal D}({\overline \Psi}_{muon})\Bigl[...\Bigr]$ 
are strongly correlated due to (9.6a,b).

Doubling of the electron and muon fields (9.6a,b)  can be examined within the 
5D invariant time theories \cite{Fanchi,Land} which allow to construct the 
independent 5D  fields $\Upsilon_{el}(x,x_5)$ and $\Upsilon_{muon}(x,x_5)$.
If the parts of the $\Upsilon_{el}(q,q^2_5)$ and $\Upsilon_{muon}(q,q_5^5)$ 
are not the same as in (9.6a,b), then they are dubled and according to
the present approach appear the fields
$(\Upsilon_{el})_-(x,x_5)$ and 
$(\Upsilon_{muon})_-(x,x_5)$ with the negative or imaginary  mass.


Nowadays  unification of the different lepton families is performed
within the 5D grand unification models \cite{Fukuyama}.
In these models the fifth dimension  is the indivisible part of the 
particle  interaction, the space-time is not asymptotically flat, 
in the extra dimensions enter other than the gravitation fields and    
is argued the  breakdown of the gauge coupling unification.  
The present formulation can be used for the 4D projections'
of the corresponding 5D electron and muon fields
$\Upsilon_{el}$ and $\Upsilon_{muon}$.
For this aim  we  put the Fourier conjugate of the complete 5D fields
$\Upsilon_{el}$ and $\Upsilon_{muon}$ on the hyperboloids 
$q^2\pm q_5^2=\pm M^2$ using the replacement of the integrals
$\int d^4q dq^2_5\{...\}$ with  
$\int d^4q dq^2_5\Bigl[\delta(q^2+ q_5^2- M^2)
\{...\}\pm \delta(q^2- q_5^2+ M^2)\{...\}\Bigr]$.
Then in analogy with (I.4a,b) and (I.7)  we get the 5D
fields $(\psi_{el})_{\pm}(x,x_5)$ and $(\psi_{muon})_{\pm}(x,x_5)$.
In the low energy region, where the asymptotically flat space is assumed,
$(\psi_{el})_{\pm}(x,x_5=0)$ and $(\psi_{muon})_{\pm}(x,x_5=0)$ determine the 
4D fields $(\Psi_{el})_{\pm}(x)$ and $(\Psi_{muon})_{\pm}(x)$. 
The gauge invariant 4D fields $(\Psi_{el})_{+}(x)$ and $(\Psi_{el})_{-}(x)$ consist
from the same four parts  as well as  $(\Psi_{muon})_{+}(x)$ and $(\Psi_{muon})_{-}(x)$.
Thus in addition to   $\Psi_{el}(x)$
 $\Psi_{muon}(x)$ we get two other fields  
$(\Psi_{el})_{-}(x)$ and $(\Psi_{muon})_{-}(x)$ with the negative 
or imaginary mass.



\vspace{0.15cm}

\centerline{\bf{10.\ Gauge transformation as generalized  
translation }}

\vspace{0.15cm}

\par 

The generalized translation 
of the four-momentum ${q^{\mu}}'=q^{\mu}-eA^{\mu}(q)$
in the equation of motion can be performed through
the gauge transformations.
In particular, the  4D and 5D gauge transformations (7.11a,b) and 
(7.8a,b)  generates 
the corresponding translations  of  the four momentum
$q^{\mu}$ and $i\partial/\partial x^{\mu}$

$${q^{\mu}}'=q^{\mu}-e{ A}^{\mu}_{\pm}(q,q_5^2)\Longleftrightarrow 
i{{\partial}\over {\partial {x'}^{\mu}}}=
i{{\partial}\over {\partial {x}^{\mu}}}-
e{ A}^{\mu}_{\pm}(x,x_5)\eqno(10.1a)$$

$${q^{5}}'=q^{5}-e{ A}^{5}_{\pm}(q,q_5^2)\Longleftrightarrow 
i{{\partial}\over {\partial {x'}^{5}}}=
i{{\partial}\over {\partial {x}^{5}}}-
e{ A}^{5}_{\pm}(x,x_5)\eqno(10.1b)$$

with the 5D  fields  ${ A}^{\mu}_{\pm}$,
 ${ A}^{5}_{\pm}$, $\psi_{\pm}$ and $\psi_{\pm}'$.

One can construct ${ A}^{\mu}_{\pm}$ and
${ A}^{5}_{\pm}$ starting from the 6D gauge translations

$${\kappa_{C}}'={\kappa_{C}}-ea_{C}(\kappa)\Longleftrightarrow 
i{{\partial}\over{\partial {\xi'}^{C}}}=i{{\partial}\over{\partial {\xi}^{C}}}-
ea_{C}(\xi)\eqno(10.2a)$$

$$\psi'(\xi)=\exp{\Bigl(ie\lambda(\xi)\Bigr)}\psi(\xi);\ \ \ 
\ \ \ \ \ \ \ \ \ \ \ \ a_{C}(\xi)=\exp{\Bigl(-ie\lambda(\xi)\Bigr)}
{{\partial}\over{\partial {\xi}^{C}}}\exp{\Bigl(ie\lambda(\xi)\Bigr)}
,\eqno(10.2b)$$
where $C\equiv \mu;5,6=0,1,2,3;5,6$ and the Fourier conjugate 
of $\psi'(\xi)$,  $\psi(\xi)$, $\lambda(\xi)$ and $a_{C}(\xi)$
are not placed on the cone $\kappa_C\kappa^C=0$.

According to invariance of the 6D cone $\kappa_C{\kappa}^C=0$ we have

$$\kappa_C{\kappa}^C=\kappa'_C{\kappa'}^C=
\Bigl(\kappa_C-ea_C(\kappa)\Bigr)\Bigl(\kappa^{C}-ea^C(\kappa)\Bigr)
=0,\eqno(10.3)$$

where $\kappa_+$ is invariant under the translations as it is 
indicated in (1.2a). 
The invariance of $\kappa_+$  
under the 4D translations (10.2a)
requires that $a_{5}(k)=-a_{6}(k)$. Substituting this condition
in (10.3) we get

$$a_{5}(k)={1\over {2 M\kappa_+}}\Bigl(
-a_{\nu}(\kappa)\kappa^{\nu}-\kappa^{\nu}a_{\nu}(\kappa)+
e a_{\nu}(\kappa)a^{\nu}(\kappa)  \Bigr);
\ \ \ \ \ \ \ \ \ 
{{ \kappa_{-}'}\over {\kappa_+}}
={{\kappa_{-}} \over {\kappa_+}}-e {{ a_{5}(k)}\over {\kappa_+}}.\eqno(10.4)$$

In the considered formulation the
  4D reduction of the 6D fields $a_{\mu}(\kappa)$
and $\psi(\kappa)$ based on the intermediate projections 
of the 6D cone $\kappa_C{\kappa}^C=0$
into 5D hyperboloids $q^2\pm q_5^2=\pm M^2$ (I.3a,b) for the domains
I, III and II, IV of the Table 1.   
The corresponding 5D reductions of the 6D
fields $a_{\mu}(\kappa)$ and $\psi(\kappa)$ are 

$$\Upsilon(q,q_5^2)={{M^2}\over 2}
\int k_+^3 d\kappa_+ \theta(\kappa_+)\psi(q,q_5^2,\kappa_+),
\eqno(10.5)$$ 
where $\Upsilon(q,q_5^2)$ is Fourier conjugate of $\Upsilon(x,x_5)$
in  (7.2) and

$${\widetilde {A}}^{\mu}(q,q_5^2)={{M^2}\over 2}
\int k_+^3 d\kappa_+ \theta(\kappa_+)
a^{\mu}(q,q_5^2,\kappa_+);\ \ \ \ \ \ 
{\widetilde {A}}^{5}(q,q_5^2)={{M^2}\over 2}
\int k_+^3 d\kappa_+ \theta(\kappa_+)
a^{5}(q,q_5^2,\kappa_+).\eqno(10.6)$$


The 5D gauge fields  ${ A}^{\mu}_{\pm}(x,x_5)$  
which Fourier conjugate 
are placed on the hyperboloids  $q^2\pm q_5^2=\pm M^2$ are constructed 
through  ${\widetilde {A} }^{\mu}$

$${ A}^{\mu}_{\pm}(x,x_5)=\int d^5y {\widetilde {A}}^{\mu}(x-y,x_5-y_5)
{\cal P}_{\pm}(y,y_5),\eqno(10.7a)$$

$${ A}^{5}_{\pm}(x,x_5)=\int d^5y {\widetilde {A}}^{5}(x-y,x_5-y_5)
{\cal P}_{\pm}(y,y_5),\eqno(10.7b)$$

The relationship between the fifth gauge field
 ${ A}^{5}_{\pm}$ (10.7b) and ${ A}^{\mu}_{\pm}$ (10.7a) 
is given in (7.13) and (7.14a,b).

The 4D gauge field ${\sf A}^{\mu}_{\pm}(x)={ A}^{\mu}_{\pm}(x,0)$
are determined through ${ A}^{\mu}_{N}(q,q^2_5)$
in the domains $N=I.II,III,IV$ as


$${\sf A}^{\mu}_{\pm}(x)=\int {{ d^4q}\over  {(2\pi)^4}} \exp{(-iqx)}\Bigl[
{\sf A}^{\mu}_{I}(q)\pm {\sf A}^{\mu}_{II}(q)+{\sf A}^{\mu}_{III}(q)
 \pm{\sf A}^{\mu}_{IV}(q)\Bigr].\eqno(10.8)$$
The convolution formula (10.7) can be represented in the 4D form 

$${ A}^{\mu}_{\pm}(x,x_5)=\int d^4y {\sf A}^{\mu}_{\pm}(x-y)
{\cal P}_{+}(y,x_5).\eqno(10.9)$$
This representation of ${ A}^{\mu}_{\pm}(x,x_5)$ indicates 
that the 5D and 4D fields ${ A}^{\mu}_{\pm}(x,x_5)$ and 
${\sf A}^{\mu}_{\pm}(x)$
satisfy the same 4D equations of motion.
On the other hand ${ A}^{\mu}_{\pm}(x,x_5)$
consists of the parts placed on the hyperboloids $q^2\pm q_5^2=\pm M^2$
and satisfy the similar to (3.1)  5D conditions 

$${{\partial^2 { A}^{\mu}_{\pm}(x,x_5)}
\over{\partial x^{\mu}\partial x_{\mu}}}+
\Bigl({{\partial^2}\over{\partial x^5\partial x_5}}+M^2\Bigr)
{ A}^{\mu}_{\mp}(x,x_5)=0, \eqno(10.10)$$

The consistency condition of (10.10) and the corresponding equation of motion
have the same form as  (4.10)   for the scalar field. 
Thus the present 5D formulation allows to construct simultaneously 
the 4D fields ${\sf A}^{\mu}_{+}(x)$ and ${\sf A}^{\mu}_{-}(x)$
which consists from the same parts ${\sf A}^{\mu}_{I}$, 
$ {\sf A}^{\mu}_{II}$, ${\sf A}^{\mu}_{III}$ and $ {\sf A}^{\mu}_{IV}$.
${\sf A}^{\mu}_{+}(x)$ and ${\sf A}^{\mu}_{-}(x)$ have the same quantum numbers, 
but the different sources. For the photon field
${\sf A}^{\mu}_{+}(x)$ the role of ${\sf A}^{\mu}_{-}(x)$
can play  the Z-boson.


It must be noted that the  gauge transformations can be performed 
also for the neutral (uncharged) particles. 
Within the nonlinear $\sigma$ model \cite{Alf,Wei}
for the triplet of the neutral auxiliary  pion fields
 $\pi^{\alpha}$ ($\alpha=1,2,3$, $\pi^{\pm}= 1/2(\pi^1\pm i \pi^2)$;
 $\pi^{0}\equiv \pi^3$) is replaced with the interpolating pion field

$$\pi^{\alpha}(x)={\cal U}(x)\chi^{\alpha}(x),\eqno(10.11)$$
where in \cite{Alf,Wei}
${\cal U}(x)=\Bigl( 1+ \chi^2(x)/4f_{\pi}^2 \Bigr)^{-1}$, 
$f_{\pi}=93\ MeV$ is the pion decay constant and 
$\chi^2=\sum_{\alpha=1}^3\chi^{\alpha}\chi^{\alpha}$.
The replacement (10.11) generates the following transformations

$${{\partial }\over {\partial x_{\mu}}}\pi^{\alpha}=
{\cal U}\Bigl[{{\partial }\over {\partial x_{\mu}}}+{\cal D}^{\mu}\Bigr]
\chi^{\alpha};\ \ \  
{\cal D}^{\mu}={\cal U}^{-1}{{\partial }\over {\partial x_{\mu}}}{\cal U}
\eqno(10.12)$$
which presents the extension  of  (7.11a,b) for the
neutral fields. The transformations (10.11) and (10.12)  generalise also
the gauge transformations (1.11) for the neutral fields.

\vspace{0.15cm}


\centerline{\bf{11.\ Summary }} 

\vspace{0.15cm}
\par

The one-to-one relationship between the 4D and 5D fields and their
equations of motion, established by present article, based on the 
equivalence of the conformal transformations  of the four momentum $q_{\mu}$ (1.1a)-(1.1e)
and the 6D rotations  on the cone  $\kappa_A\kappa^A=0$ (1.2a)-(1.2d)
and its 5D projections on the two
invariant forms $q^2\pm q_5^2=\pm M^2$ of the  $O(2,3)$ and $O(1,4)$ subgroups 
of the conformal group $O(2,4)$.
The 6D cone $\kappa_A\kappa^A=0$ and its 5D projections $q^2\pm q_5^2=\pm M^2$
are invariant under the 6D rotations and the corresponding 
4D conformal transformations.
Consequently, the 4D projection of the 6D field $\varsigma(\kappa)$ with the intermediate
projection on the two hyperboloids $q^2\pm q_5^2=\pm M^2$
determine the two 5D fields $\varphi_{1}(x,x_5)$  (I.4a) and 
$\varphi_{2}(x,x_5)$ (I.4b).  The  Fourier conjugate of these 5D fields are 
placed  on the hyperboloids  $q^2\pm q_5^2=\pm M^2$ before and after conformal 
transformations. 
Consequently the 5D fields  $\varphi_{+}=\varphi_1 + \varphi_2$
and  $\varphi_{-}=\varphi_1 - \varphi_2$ (I.7)  are defined on the
whole domain of $-\infty<q^2<\infty$ 
 and  reproduce the 4D fields $\Phi_{\pm}(x)=\varphi_{\pm}(x,x_5=0)$. 
In addition the fields $\varphi_{\pm}(x,x_5)$ 
satisfy the  5D constrains (3.1) that have the form of the coupled sourceless 5D equations.
The 5D equation of motion for $\varphi_{\pm}$
(4.1a)  and their 4D reductions (4.1b)
are embedded into these sourceless coupled 5D  conditions (3.1). 
The consistency conditions of the 5D equation of motion (4.1a) 
and the  5D constrains (3.1) generate the constrains for
$\partial \varphi_{\pm}/\partial x_5$. 
The similar sourceless 5D coupling condition (7.1)  
 are  valid for the interacting  fermion fields $\psi_+$ and $\psi_-$.



This scheme can be used for the 5D extension of the 4D models and for the 
4D reductions of the 5D formulations.
The parts of the 5D field  $\phi(q,q_5^2=M^2-q^2)$ and $\phi(q,q_5^2=M^2+q^2)$
unambiguously determine the Fourier conjugate of the 4D fields $\Phi_{+}(x)$
and  $\Phi_{-}(x)$   (I.10)-(I.11). The same expressions determine also 
the Fourier conjugate of the 5D fields  $\varphi_1$ (I.4a), $\varphi_2$ (I.4b)
and $\varphi_{\pm}$ (I.7).  These parts of the single 5D field 
$\phi(q,q_5^2)$  determine unambigously the two 4D fields $\Phi_{\pm}(x)$ 
 with the same quantum numbers, but with the different masses and  sources. 
And vice versa, starting from the 4D field $\Phi_{+}(x)$
one can construct  $\Phi_{-}(x)$ and the parts of the 5D field 
$\phi(q,q_5^2=M^2-q^2)$ and $\phi(q,q_5^2=M^2+q^2)$.
This doubling of the 4D fields is result the intermediate projection
of the 6D field placed on the 6D cone $\kappa_A\kappa^A=0$  into the two 5D fields $\varphi_{\pm}$ 
embedded into two invariant forms $q^2\pm q_5^2=\pm M^2$.
The considered  intermediate 5D projections take into account the symmetry
under the inversion $q'_{\mu}=-M^2q_{\mu}/q^2$ 
and reflection ${q'}^2=-q^2$ between the domains of these forms. 
As it is mentioned in the last two paragraphs of Sect. 2,
the stereographic and other 5D projections of the 6D cone $\kappa_A\kappa^A=0$
can be reproduced through the  considered projections on the hyperboloids  
$q^2\pm q_5^2=\pm M^2$. 

The boundary conditions of the 5D fields $\varphi_{\pm}$ (I.8)  and $\psi_{\pm}$ (7.4)
at $x_5=0$ can be extended for 
an arbitrary value $x_5=t_5$ if one replaces $x_5$ by
$x_5-t_5$ in the definitions of the 5D fields (I.4a,b) and (i.7). In particular, $t_5=\sqrt{x_0^2-{\bf x}^2}$
in the formulation within  the relativistic invariant  time theories\cite{Fanchi,Land}.

 The common parts of the  4D interacted fields 
$\Phi_{+}(q)$ and $\Phi_{-}(q)$ in (I.10)-(I.11) 
allow to separate the $4!=24$  different  fields 
with the same quantum numbers and with the different masses and sources.
This  unification scheme of the interacting Heisenberg  
fields can be applied for description of the nucleons and resonances in the  $P11$ states 
$N(1440)$, $N(1710)$, ... \cite{PDG}
based on an additional dynamical mechanism for the generation of the
resonances. Similarly,  one can combine  the interacting Heisenberg
fields of the pions and the resonances
with the same quantum numbers $\pi(1300)$.

The particle states with the same quantum numbers and the different masses
and sources can be constructed  within
the various 5D relativistic invariant time theories  \cite{Fanchi,Land}.
The different 5D fields $\phi_+(q,q_5^2)$ and $\phi_-(q,q_5^2)$ in these theories 
do not have the common parts $\phi_+(q,q_5^2=M^2-q^2)$ and $\phi_-(q,q_5^2=M^2+q^2)$.
Nevertheless the present formulation requires doubling of the
 5D  fields, i.e. instead of the $\phi_+(x,x_5)$ and $\phi_-(x,x_5)$ we get
$(\phi_+)_{\pm}(x,x_5)$ and $(\phi_-)_{\pm}(x,x_5)$, 
where $(\phi_+)_{-}(x,x_5)$ and $(\phi_-)_{-}(x,x_5)$ must have the negative or imaginary masses.
The relationships between the different masses and sources for the 5D fields
$(\phi_+)_{\pm}(x,x_5)$ and $(\phi_-)_{\pm}(x,x_5)$ are determined 
by constrains for  the   $\partial (\phi_+)_{\pm}/partial x_5$ and 
$\partial (\phi_-)_{\pm}/partial x_5$.
Thus the present approach allows to get 
the 4D projections of the 5D fields from \cite{Fanchi,Land} for $x_5=0$. 

It must be noted, that if the 4D fields $\Phi_{\pm}(x)$ are determined through the 
parts of the 5D fields   
$\phi_{+}(q,q^2_5=M^2\mp q^2)$  and $\phi_{-}(q,q^2_5=M^2\mp q^2)$, 
then the Fourier conjugate of the observed 4D fields $\Phi_{\pm}(q)$ can have
the jump  singularities at $q^2=0,\pm M^2$.


Presently the different lepton families including the electron and muon fields
$\Upsilon_{el}$ and $\Upsilon_{muon}$ are constructed within the 5D grand unification theories 
\cite{Fukuyama,Fujimoto}. In this formulation the 5D electron and muon fields
$\Upsilon_{el}(q,q_5^2=M^2\mp q^2)$ and $\Upsilon_{muon}(q,q_5^2=M^2\mp q^2)$ are independent. 
The present scheme of the 4D reduction of the 5D fields allows to get the 4D projections
of these 5D fields with the doubling of the 4D electron and muon fields   
$(\Psi_{el})_{\pm}(x)$ and $(\Psi_{muon})_{\pm}(x)$, where  $(\Psi_{el})_{-}$ and $(\Psi_{muon})_{-}$,
 have the negative or imanary masses.
These 4D projections of the 5D fields restore the 
4D gauge invariance, because the gauge transformations are considered as the
generalized 4D translation in the momentum space  (see Sect. 10) and
translations of the four momentum preserve  invariance of the 
6D cone $\kappa_A\kappa^A=0$ and their projections on the 5D invariant forms.


Other kind of the unification of the  4D electron and muon fields in this approach is considered in 
the Section 9 for the Standard Model \cite{Weinberg},  
where $\Phi_{el}(q)$ and $\Phi_{muon}(q)$ consist of the same parts (9.6a,b)
of the complete 5D field $\Upsilon(q,q^2_5=M^2\mp q^2)$.
The coupling between the electron and muon fields are strongly suppressed
in the Standard Model. Therefore, 
 the common structure of the electron and muon fields can not be observed
in the perturbation series of the corresponding 4D equations.




\vspace{0.5cm}

I thank V.G.Kadyshevsky for numerous helpful discussions. 
I am thankful to  A. Machavariani (junior)
and G. M\"unster for the current interest in this work.




\vspace{0.15cm}

\begin{thebibliography}{99}
\bibitem{Appelquist} T. Appelquist, A. Chodos and P.G.O. Freund. 
Modern Kaluza-Klein Theories, Addison-Wesley, Monte-Park, 1987.
\bibitem{Bailin} D. Bailin and A. Love.Rep. Prog. Phys. {\bf 50} (1987) 1087.
\bibitem{Wesson} P. S. Wesson. Space- Time - Matter; Modern Kaluza-Klein 
Theory, World Scientific, 2000.
\bibitem{Cian} F. Cianfrani, A. Marrocco and G. Montani. 
Int. Jour. Modern Phys. {\bf D14} (2005) 1195.
\bibitem{Fukuyama} T. Fukuyama, 
Int. Jour. Mod. Phys. {\bf A28} (2013) 1330008.
\bibitem{Fujimoto} T. Fujimoto, T. Nagasawa. K Nishiwaki and M. 
Sakamoto. Prog. Theor. Exp. Phys. {\bf B07} (2013) 023. 


\bibitem{IZ} C. Itzykson and J. B. Zuber.
Quantum Field Theory; New York, McGrew-Hill, 1980.

\bibitem{Fanchi} J. L. Fanchi, Found. Phys.{\bf 22}(1993)487.
\bibitem{Land} M. C. Land, Found. Phys.{\bf 27}(1997) 19.




\bibitem{Dir} P.A.M. Dirac, Ann. Math. {\bf 37} (1936) 429.
\bibitem{Kas} H. A. Kastrup, Phys. Rev.{\bf 150}(1966) 1183;
\bibitem{Ca1} L. Castell, Nuovo Cim.{\bf A46}(1966) 1; 
\ L. Castell, Nuovo Cim.{\bf A49}(1967) 285.
\bibitem{Ca2} L. Castell, Nucl. Phys. {\bf B4}(1967) 343.


\bibitem{M1} G. Mack and A. Salam, Ann. Phys.{\bf 53}(1969) 174.
\bibitem{Gatto} S. Ferrara, R. Gatto and A.F. Grilo, Ann. Phys.
{\bf 76} (1973) 161;
S. Ferrara, R. Gatto and A.F. Grilo,in
 Scale and Conformal Symmetry in Hadron
Physics (ed. R. Gatto), Wiley, New York, 1974;
\bibitem{Alf} V. De Alfaro, S.  Fubini, G.  Furlan and C.  Rosseti,
Currents in Hadron Physics; (North-Holland, Amsterdam) 1973.
\bibitem{Kon} B.G. Konopelchenko,Sov. J. Elem. Part. and At. Nucl. 
(in Russian) {\bf 11}(1977) 135.
\bibitem{BK} J. Beckers, J. Harnad, M. Perroud and P. Winternitz,
J. Math. Phys.{\bf 19}(1978) 2126.
\bibitem{BaR} A.O.Barut and R. Raczka, Theory of Group Representations
and Applications;PWN,Warszawa, 1977.
\bibitem{BR} P. Budinich and R. Raczka, Found. Phys.{\bf 23}(1993) 599.
\bibitem{Mirman} R. Mirman, 
Quantum Field Theory Conformal Group Theory Conformal Field Theory: 
Mathematical And Conceptual Foundations Physical And Geometrical 
Applications;Paperback, Backinprint.com, 2005.   
\bibitem{Braun} V.M. Braun, G. P. Korchemsky  and D. M\"uller,
 Prog. Part. Nucl. Phys.  {\bf 51} (2003) 311.
\bibitem{Fradkin} E. S. Fradkin  and M.Y. Palchik,
Conformal\ Quantum\ Field\ Theory\ in \ D-Dimensions,
in: Mathematics\ and\ its\ Applications\ V.376\ Kluwer,Dordrecht,
 Netherlands,\ New\ York,\ 1996; E. S. Fradkin  and M.Y. Palchik,
Phys.Rep.{\bf 300} (1998) 1.
\bibitem{Todorov}  I. T. Todorov, Conformal\ Description\ of\
Spinning\ Particles;\ Springer,\ New\ York,\ 1986;
 Todorov I. T., Minchev M.C. and Petkova V.B. //
Conformal\ covariance\ in\ quantum\ field\ theory; (Scuola
\ Normale\ Superiore,\ Pisa,\ 1978.




\bibitem{Budinich} P. Budinich,  Found. Phys. {\bf 32}(2002) 1347.
\bibitem{Bacr} H. Bacry, Ann. Ins. H. Pouncare, {\bf 49}(1988) 245;
H. Bacry, Localizabelity\ and\ Space\ in\ Quantum\ Physics;\ Lect.
Notes\ in\ Phys.\ {\bf 308}
(Springer,\ Berlin\ Heidelberg,\ 1988).

\bibitem{K1} V. G. Kadyshevsky, J.Exp. Theor. Phys. 
(in Russian){\bf 41}(1961) 1885.
\bibitem{K2} V. G. Kadyshevsky, 
Sov.J. Elem. Part. and At. Nucl. (in Russian) {\bf 11}(1980) 5;
Preprint JINR,(in Russian) ░2-84-753.
\bibitem{K3} V. G. Kadyshevsky and M. D. Mateev, Nuovo Cim.{\bf 87A}(1985)324;
M. V. Chizhov, A. D. Donkov, R. M. Ibadov, V. G. Kadyshevsky and M. D. Mateev,
Nuovo Cim.{\bf 87A}(1985)351 and 373.
\bibitem{Heisenberg} W. Heisenberg, Ann. Phys. (Leipzig){\bf 5}(1938) 20;
H. P. D\"urr and  W. Heisenberg, Z. Natur. {\bf 16a}(1961) 726.
\bibitem{Markov} M. A. Markov, Suppl. Prog. Theor. Phys.
Commamemory Issue for 30-th Anniveversary of Meson Theory by Dr. A. Yukawa
(1965) 865; J.Exp. Theor. Phys. (in Russian) {\bf 51}(1966) 878.

\bibitem{prepr} A.I.Machavariani, Preprint 
arXiv/math-ph/0611083v1 (2006) which is the renewed version of the 
lectures presented\ in \ arXiv/hep-th/0504030 (2005).



\bibitem{M2} G. Mack and I. T. Todorov, Phys. Rev.{\bf D8}(1973) 1764.
\bibitem{BD} J. R. Bjorken and S. D. Drell. Relativistic Quantum Fields;
New York, McGrew-Hill, 1963.
\bibitem{Weinberg} S. Weinberg,
The Quantum  Theory of Fields; Cambridge, University Press, 1995 and 1996.


\bibitem{Wei} S. Weinberg, Phys. Rev.{\bf D2}(1970)674; ibid
{\bf 177}(1969)2604.

\bibitem{PDG}K. Nakamura et al.(Particle Data Group) J. Phys. G {\bf 37}
 (2010) 075021.



\end{thebibliography}
\end{document}